\documentclass{aastex62}

\usepackage{graphicx}   
\usepackage{amsmath}    
\usepackage{amssymb}    
\usepackage{amsfonts}
\usepackage{epsfig,rotating}
\usepackage[T1]{fontenc}
\usepackage{ae,aecompl}
\usepackage{subfigure}
\usepackage{longtable}
\usepackage{booktabs}
\usepackage{newtxtext}
\usepackage[varvw]{newtxmath}
\usepackage{ulem}

\def \kms  {$\rm km\,s^{-1}$}

\def\purple#1 {{\textcolor{purple}{#1}}\ }
\def\red#1 {\textcolor{red}{#1}}
\def\new#1 {{\bf #1 }}
\def\blue#1 {{\textcolor{blue}{#1}}\ }

\def \Ctw {$^{12}$C}
\def \Cth {$^{13}$C}

\graphicspath{{./}{figures/}}

\accepted{31 May, 2022}

\shorttitle{Dense gas in  M$\,$82}
\shortauthors{Li et al.}

\begin{document}

\title{ Properties of dense molecular gas along the major axis of M$\,$82}

\email{lifei.astro@gmail.com;zzhang@nju.edu.cn}


\author[0000-0003-3353-7106]{Fei Li}
\affiliation{School of Astronomy and Space Science, Nanjing University, Nanjing 210093, People’s Republic of China\\}
\affiliation{Key Laboratory of Modern Astronomy and Astrophysics (Nanjing University), Ministry of Education, Nanjing 210093, People’s Republic of China\\}
\affiliation{Shanghai Astronomical Observatory, Chinese Academy of Sciences, 80 Nandan Road, Shanghai 200030, People’s Republic of China\\}

\author[0000-0002-7299-2876]{Zhi-Yu Zhang}
\affiliation{School of Astronomy and Space Science, Nanjing University, Nanjing 210093, People’s Republic of China\\}
\affiliation{Key Laboratory of Modern Astronomy and Astrophysics (Nanjing University), Ministry of Education, Nanjing 210093, People’s Republic of China\\}

\author[0000-0001-6106-1171]{Junzhi Wang}
\affiliation{Shanghai Astronomical Observatory, Chinese Academy of Sciences, 80 Nandan Road, Shanghai 200030, People’s Republic of China\\}
\affiliation{School of Physical Science and Technology, Guangxi University, Nanning 530004, People’s Republic of China \\}

\author[0000-0002-2581-9114]{Feng Gao}
\affiliation{Hamburger Sternwarte, Universitaet Hamburg, Gojenbergsweg 112, 21029, Hamburg, Germany\\}

\author[0000-0003-1275-5251]{Shanghuo Li}
\affiliation{Korea Astronomy and Space Science Institute, 776 Daedeokdae-ro, Yuseong-gu, Daejeon 34055, Republic of Korea\\}

\author[0000-0002-0818-1745]{Jing Zhou}
\affiliation{School of Astronomy and Space Science, Nanjing University, Nanjing 210093, People’s Republic of China\\}
\affiliation{Key Laboratory of Modern Astronomy and Astrophysics (Nanjing University), Ministry of Education, Nanjing 210093, People’s Republic of China\\}

\author[0000-0003-0477-1606]{Yichen Sun}
\affiliation{School of Astronomy and Space Science, Nanjing University, Nanjing 210093, People’s Republic of China\\}
\affiliation{Key Laboratory of Modern Astronomy and Astrophysics (Nanjing University), Ministry of Education, Nanjing 210093, People’s Republic of China\\}

\author[0000-0002-7532-1496]{Ziyi Guo}
\affiliation{School of Astronomy and Space Science, Nanjing University, Nanjing 210093, People’s Republic of China\\}
\affiliation{Key Laboratory of Modern Astronomy and Astrophysics (Nanjing University), Ministry of Education, Nanjing 210093, People’s Republic of China\\}

\author[0000-0001-6016-5550]{Shu Liu}
\affiliation{CAS Key Laboratory of FAST, National Astronomical Observatories, Chinese Academy of Sciences, Beijing 100012, China}

\begin{abstract}

Dense gas is important for galaxy evolution and star formation.  Optically-thin
dense-gas tracers, such as isotopologues of HCN, HCO$^+$, etc., are very
helpful to diagnose excitation conditions of dense molecular gas. However, previous
studies of optically-thin dense-gas tracers were mostly focusing on average
properties of galaxies as a whole, due to limited sensitivity and angular
resolution. M$\,$82, a nearby prototype starburst galaxy, offers a unique case
for spatially-resolved studies with single-dish telescopes. With the IRAM 30-m
telescope, we observed the $J = 1\rightarrow0$ transition of H$^{13}$CN,
HC$^{15}$N, H$^{13}$CO$^+$, HN$^{13}$C, H$^{15}$NC, and SiO $J =
2\rightarrow1$, HC$_3$N $J = 10\rightarrow9$, H$_2$CO $J = 2\rightarrow1$
toward five positions along the major axis of M$\,$82.  The intensity ratios of
$I$(HCN)/$I$(H$^{13}$CN) and $I$(HCO$^+$)/$I$(H$^{13}$CO$^+$) show a
significant spatial variation along the major axis,  with lower values in the
central region than those on the disk, indicating higher optical depths in the
central region. The optical depths of HCO$^+$ lines are found to be systematically
higher than those of HCN lines at all positions. Futhermore, we find that the
$^{14}$N/$^{15}$N ratios have an increasing gradient from the center to the
outer disk.  
\end{abstract}



\section{Introduction} 

Observations show that star-formation activities are closely connected with
dense molecular gas both in the Milky Way and in other galaxies.
Heiderman2010,Lada2010,Lada2012. The dense molecular gas
are directly involved into star formation activities, and it can be probed with
molecular lines with critical densities ($n_{\rm crit}$) greater than
$10^4$ $\rm{cm}^{-3}$, such as multi-$J$ transitions of HCN and HCO$^+$
\citep{Lada1992,Kohno1999,Kennicutt2012}. With observations of HCN $J =
1\rightarrow0$ toward 65 galaxies, \cite{Gao2004a} found a tight linear
correlation between the luminosities of HCN $J$ = 1$\rightarrow0$ and total
infrared emission. However, most dense gas tracers are normally optically thick
both in Galactic giant molecular clouds (GMCs) and in external galaxies
\citep{Wang2014,Meier2015,Jimenez2017,Li2017,Li2020}. Therefore, there is a large
uncertainty in estimating the dense gas mass from a single transition line of a high dipole moment molecule, which is similar to the issue
of the CO-to-H2 conversion factors \citep{Narayanan2012,Bolatto2013}.

Optically thin dense-gas tracers, such as the isotopologues of HCN and
HCO$^+$, could help better revealing dense gas properties in galaxies, such as volume
density, temperature, and excitation conditions. One can determine optical
depths of dense gas tracers, using their isotopologue line ratios
\citep{Henkel1998,Martin2010,Wang2014,Li2020}.  Because of their low abundances, isotopologue lines
are mostly optically thin, and thus they can be used to accurately determine dense
molecular gas properties and help study dense gas--star formation relations in
different galaxies \citep{Li2020}.

In most galaxies, however, isotopologue lines of dense gas tracers are too
faint to be detected. Only a few detections of such lines have been reported in
nearby galaxies \citep{Henkel1998,Wang2014,Wang2016,Tunnard2015,Li2020}, which
either are limited within only the central regions of galaxies, or take global
properties of galaxies as a whole \citep{Wang2014,Li2020}.  

Toward one of the nearest starburst galaxy, M$\,$82, \cite{Li2020} observed
multi-$J$ HCN lines in the central region. Intriguingly, they found that the
line profiles of H$^{13}$CN $J$=1$\rightarrow$0 and $J$=3$\rightarrow$2 are not
consistent with each other. The $J$=1$\rightarrow$0 and $J$=3$\rightarrow$2
transitions of H$^{13}$CN are dominated by the red-shifted and the blue-shifted
velocity components, respectively.  Because H$^{13}$CN $J$=3$\rightarrow$2 has
a higher upper energy level and a higher critical density than those of
H$^{13}$CN $J$=1$\rightarrow$0, such a difference may indicate
velocity-dependent excitation conditions, isotope abundance variations, or
exotic radiative transfer processes. It is natural to rise the following
questions: How much would dense gas properties vary at different locations and
velocity components of a galaxy? How would dense gas isotopologue abundances
change along the galactic disk? What physical mechanisms would regulate dense
gas properties on galactic scales?  

To address these questions above, the best way is to obtain spatially resolved dense
gas isotopologue measurements, with deep integration. In this paper, we present
new IRAM 30-m observations of the $^{13}$C and $^{15}$N isotopologues of HCN,
HCO$^+$, and HNC, along the major axis of M$\,$82. We describe observations and
data reduction in Section \ref{sec:obs}. The new spectra, their intensities,
optical depths, and abundance ratios are presented in Section
\ref{sec:results}. Section \ref{sec:discussion} discussed possible mechanisms
that may dominant or bias these line ratios. Final conclusions are summarized in
Section \ref{sec:summary}.


\section{Observations and data reduction}
\label{sec:obs}


Observations were performed with the IRAM 30-m telescope, at Pico Veleta, Spain
during February 2019 (Project number: 186-18, PI: Feng Gao). A total number of four 
different pointings were used to sample the off-center region along the major axis of M82, 
with a typical pointing offset of 15$^{\prime\prime}$. We list these pointing positions in Table 
\ref{table:M82_position} below. Figure
\ref{fig:M82_cor} shows the observed four off-center positions overlaid on the
velocity-integrated flux (moment 0) map of HCN \citep{Salas2014}, where
positions-1 and -2 are located in the South-Western side of M$\,$82 while
positions-4 and -5 are located in the North-Eastern side \citep{Aladro2011b}.

The Eight Mixer Receiver (EMIR) with dual-polarization and the Fourier
Transform Spectrometers (FTS) backend with a 8-GHz frequency coverage for each
band and a 195-kHz spectral resolution were used. EMIR was configured  to the
mutual observing mode for both E-90 (at 3 mm) and E-150 (at 2 mm) receivers.
To verify that the signal is from the sky frequency rather than radio-frequency
interference (RFI) from the Earth or from the backend, two different local
oscillator (LO) tuning setups were used during the observations (see Table
\ref{tab:parameter_obs}). 

All observations were performed with the wobbler switching mode, which has a
beam throw distance of $\pm$60$^{\prime\prime}$ and a switching frequency of
0.5 Hz. Telescope pointing was checked every two hours with nearby strong quasi-stellar
objects, while focus was checked at the beginning of each observation.

The beam sizes of the IRAM 30-m telescope are approximately 29$^{\prime\prime}$
and 16$^{\prime\prime}$ at 87 GHz and 145 GHz, respectively. Typical system
temperature are $\le$ 100 K and $\sim$ 120 K, at 3-mm and 2-mm band,
respectively. The antenna temperature ($T_{\rm A}^{*}$) is converted to the
main beam temperature ($T_{\rm mb}$), using $T_{\rm mb}$=$T_{\rm A}^{*}\times
F_{\rm eff}$/$B_{\rm eff}$, where the forward efficiencies ($F_{\rm eff}$) are
0.95 and 0.93, beam efficiencies ($B_{\rm eff}$) are 0.81 and 0.74, at 3-mm and
2-mm band, respectively. Each spectrum was read out every 2 minutes. The
on-source time ranges from 3.5 to 4 hours towards each position.


\begin{figure*}
\centering
\includegraphics[scale=0.30]{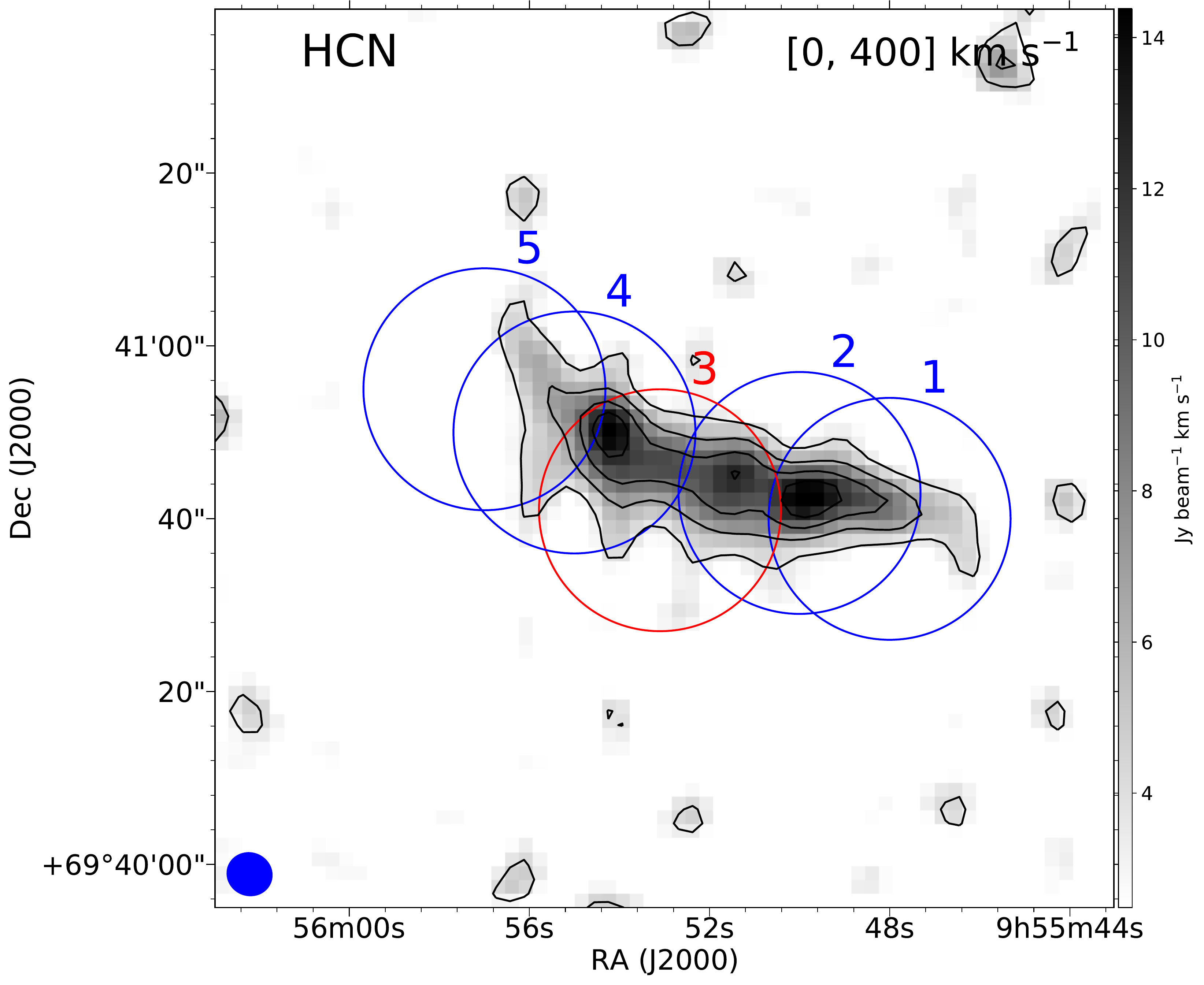}
\vspace{2em}
\caption{Velocity-integrated intensity (moment 0) maps of HCN
        $J$=1$\rightarrow$0 emission from \citet{Salas2014} using CARMA. The
        synthesis beam of CARMA data has a FWHM of $\sim$6 $^{\prime\prime}$,
        which is shown in the lower left corner.  IRAM 30-m beam (29 $''$)
        coverage are shown in blue (off-centre) and red (central position;
        \cite{Li2020}) circles. The contours start from 3 $\sigma$ to 6
        $\sigma$, with an 1-$\sigma$ level of 2.5 K\,km\,s$^{-1}$.}

\label{fig:M82_cor}
\end{figure*}


We adopt GILDAS/CLASS\footnote{http://www.iram.fr/IRAMFR/GILDAS} \citep[GILDAS
team 2013]{Pety2005} to reduce all spectral data as the following order: First,
we check all spectra by eye and removed questionable ones. Second, for each
position we average all reliable spectra into one spectrum. We then fit a
first-order polynomial baseline and subtract it from the averaged spectrum.
Last, the baseline-subtracted spectra are smoothed to a velocity resolution of
$\sim$ 35 km s$^{-1}$.



\begin{table}
\scriptsize
    \begin{center}
      \caption{Observed positions along the major axis of M$\,$82}\label{tab:source}
         \label{table:M82_position}
      \begin{tabular}{lllllllll}
      \\
    \hline
    \hline
Position           & R.A.       & Dec.     & $R_{\rm GC}$  \\
                   & (J2000)    & (J2000)  & kpc       \\
            \hline
1                  & 09:55:48   & 69:40:40 & 0.45  \\
2                  & 09:55:50   & 69:40:43 & 0.28\\
3(center)          & 09:55:53.1 & 69:40:41 & 0\\
4                  & 09:55:55   & 69:40:50 & 0.23\\
5                  & 09:55:57   & 69:40:55 & 0.42\\

    \hline
      \end{tabular}

  \end{center}
Positions-1 and -2 locate in the SW lobe. Positions-4 and -5 locate in the NE lobe.
$R_{\rm GC}$ is the distance from the Galactic Centre.

\end{table}

\begin{table*}
\centering

      \caption{The basic properties of these five positions}
      
       \label{tab:parameter_obs}
      \begin{minipage}{160mm}
      \footnotesize
        \begin{tabular}{p{1.2cm}p{2cm}p{1.6cm}p{1.6cm}p{1.6cm}p{1.8cm}p{1.8cm}}
        
             \hline
            \hline
            
      {Position}                 & {Observing dates} & {$t_{\rm on\,source}$} & {$T^{\rm 3 mm}_{\rm sys}$} & {$T^{\rm 2 mm}_{\rm sys}$} & {LO(3 mm) } & {LO(2 mm)$^{~\rm a}$ } \\
                                 & YYYY-MM-DD        & (min)                  & (K)                        & (K)                        & (GHz)       & (GHz)   \\

             \hline

Position-1                       & 2019-Feb-06       & 47                     & 83                         & 120                        & 86.34       & 146.97 \\
                                 & 2019-Feb-07       & 20                     & 93                         & 130                        & 86.29       & 146.92 \\
                                 & 2019-Feb-10       & 48                     & 110                        & $-$                        & 86.15       & $-$ \\
Position-2                       & 2019-Feb-06       & 26                     & 93                         & 120                        & 86.34       & 146.97 \\
                                 & 2019-Feb-07       & 25                     & 94                         & 127                        & 86.29       & 146.92 \\
                                 & 2019-Feb-08       & 39                     & 105                        & $-$                        & 86.24       & $-$ \\
                                 & 2019-Feb-09       & 26                     & 106                        & $-$                        & 86.19       & $-$ \\
Position-4                       & 2019-Feb-06       & 22                     & 96                         & 134                        & 86.34       & 146.97 \\
                                 & 2019-Feb-07       & 25                     & 90                         & 121                        & 86.29       & 146.92 \\
                                 & 2019-Feb-08       & 45                     & 109                        & $-$                        & 86.24       & $-$ \\
                                 & 2019-Feb-09       & 26                     & 104                        & $-$                        & 86.19       & $-$ \\
Position-5                       & 2019-Feb-06       & 16                     & 101                        & 142                        & 86.34       & 146.97 \\
                                 & 2019-Feb-07       & 35                     & 92                         & 125                        & 86.29       & 146.92 \\
                                 & 2019-Feb-10       & 38                     & 116                        & $-$                        & 86.15       & $-$ \\
\hline  
 
        \end{tabular}\\

  \begin{list}{}{}
\item[${\mathrm{a}}$]Tuning frequencies of the Local oscillator (LO) during the observations.
\end{list}

 \end{minipage}
 \end{table*}

\begin{table*}
\begin{center}

\caption{Integrated Intensities at All Positions}    
\label{tab:all_position}
\scriptsize
\begin{tabular}{p{16.3mm}p{11mm}lllllllllllllllllllll}
\hline
\hline
                                       &                     & \multicolumn{10}{c}{$I$(K km s$^{-1}$)}  \\
 \cline{3-13}
{Line}                                 & {$\nu _{\rm rest}$} & {Position-1}                              & \multicolumn{2}{c}{Position-2} &                             & \multicolumn{2}{c}{Position-3} &                           & \multicolumn{2}{c}{Position-4 } &  & \multicolumn{2}{c}{Position-5 }\\
 \cline{4-5} \cline{7-8} \cline{10-11}
                                        & {(GHz)} & Blue-shifted         & Blue-shifted         & Red-shifted                 &  & Blue-shifted              & Red-shifted          &  & Blue-shifted            & Red-shifted         &  & Red-shifted    \\
\hline

HC$^{15}$N 1-0     & 86.055  & $0.08\pm0.025$ & $0.29\pm0.04$  & $<0.11$       &  & $0.29\pm0.02$  & $0.24\pm0.02$  &  & $-     $      & $<0.08      $  &  & $<0.08$  \\
H$^{13}$CN 1-0     & 86.340  & $0.23\pm0.05$  & $0.38\pm0.04$  & $0.27\pm0.04$ &  & $0.21\pm0.04$  & $0.33\pm0.04$  &  & $<0.15$       & $0.30\pm0.05$  &  & $0.14\pm0.02$ \\
SiO 2-1            & 86.847  & $<0.20$        & $<0.12$        & $-$           &  & $0.17\pm0.04$  & $0.25\pm0.04$  &  & $-     $      & $0.25\pm0.07$  &  & $0.29\pm0.07$ \\
H$^{13}$CO$^+$ 1-0 & 86.754  & $0.53\pm0.07$  & $0.80\pm0.04$  & $0.12$        &  & $0.50\pm0.04$  & $0.71\pm0.04$  &  & $0.26\pm0.07$ & $0.68\pm0.07$  &  & $0.36\pm0.07$ \\
HCO 1-0            & 86.670  & $0.45\pm0.07$  & $0.46\pm0.04$  & $-$           &  & $0.27\pm0.04$  & $0.34\pm0.04$  &  & $-     $      & $-$            &  & $<0.20      $  \\
HN$^{13}$C 1-0     & 87.091  & $<0.10$        & $<0.12$        & $-$           &  & $<0.08      $  & $0.13\pm0.03$  &  & $-     $      & $0.15\pm0.05$  &  & $0.16\pm0.05$  \\
HCN 1-0            & 88.632  & $14.46\pm0.26$ & $19.63\pm0.04$ & $6.04\pm0.04$ &  & $11.57\pm0.04$ & $15.47\pm0.04$ &  & $3.71\pm0.03$ & $21.09\pm0.03$ &  & $14.26\pm0.06$ \\
H$^{15}$NC 1-0     & 88.866  & $<0.09$        & $<0.12$        & $-$           &  & $-           $ & $<0.15$        &  & $-          $ & $<0.10       $ &  & $<0.18      $  \\
HCO$^+$     1-0    & 89.189  & $22.29\pm0.04$ & $30.23\pm0.04$ & $8.14\pm0.04$ &  & $18.16\pm0.05$ & $22.64\pm0.05$ &  & $6.18\pm0.04$ & $29.89\pm0.04$ &  & $19.69\pm0.05$ \\
HNC 1-0            & 90.663  & $6.60\pm0.03$  & $9.67\pm0.05$  & $2.45\pm0.05$ &  & $6.00\pm0.05$  & $7.13\pm0.05$  &  & $1.74\pm0.03$ & $9.68\pm0.03 $ &  & $5.82\pm0.05 $ \\
HC$_3$N 10-9       & 90.979  & $0.64\pm0.03$  & $1.19\pm0.04$  & $0.40\pm0.04$ &  & $0.65\pm0.03$  & $1.06\pm0.03$  &  & $0.11\pm0.03$ & $1.14\pm0.03 $ &  & $0.61\pm0.04 $\\
H$_2$CO  2-1       & 145.603 & $1.73\pm0.06$  & $2.86\pm0.08$  & $-$           &  & $0.67\pm0.07$  & $2.35\pm0.07$  &  & $0.17\pm0.05$ & $3.22\pm0.05 $ &  & $1.40\pm0.06 $ \\

             \hline

\end{tabular}
\end{center}
Notes: Velocity-integrated intensities are calculated with fixed velocity ranges of
0--200 km s$^{-1}$ and 200--400 \kms\ for the blue-shifted and the red-shifted
component, respectively.  Errors of the velocity-integrated intensities are
calculated with $\Sigma (I)$ = $\sigma_{\rm line-free}^{\rm chan} \times
\sqrt{\Delta V \delta \varv}$, where  $\sigma_{\rm line-free}^{\rm chan}$ is
the r.m.s. noise level obtained from the line-free channels at the
corresponding velocity resolution;   $\delta v$ is the velocity resolution (36
\kms); $\Delta V$ is the line width (200 km s$^{-1}$). 

\end{table*}

\section{results}
\label{sec:results}

\begin{figure*}
\centering
\subfigure {\includegraphics[scale=0.24]{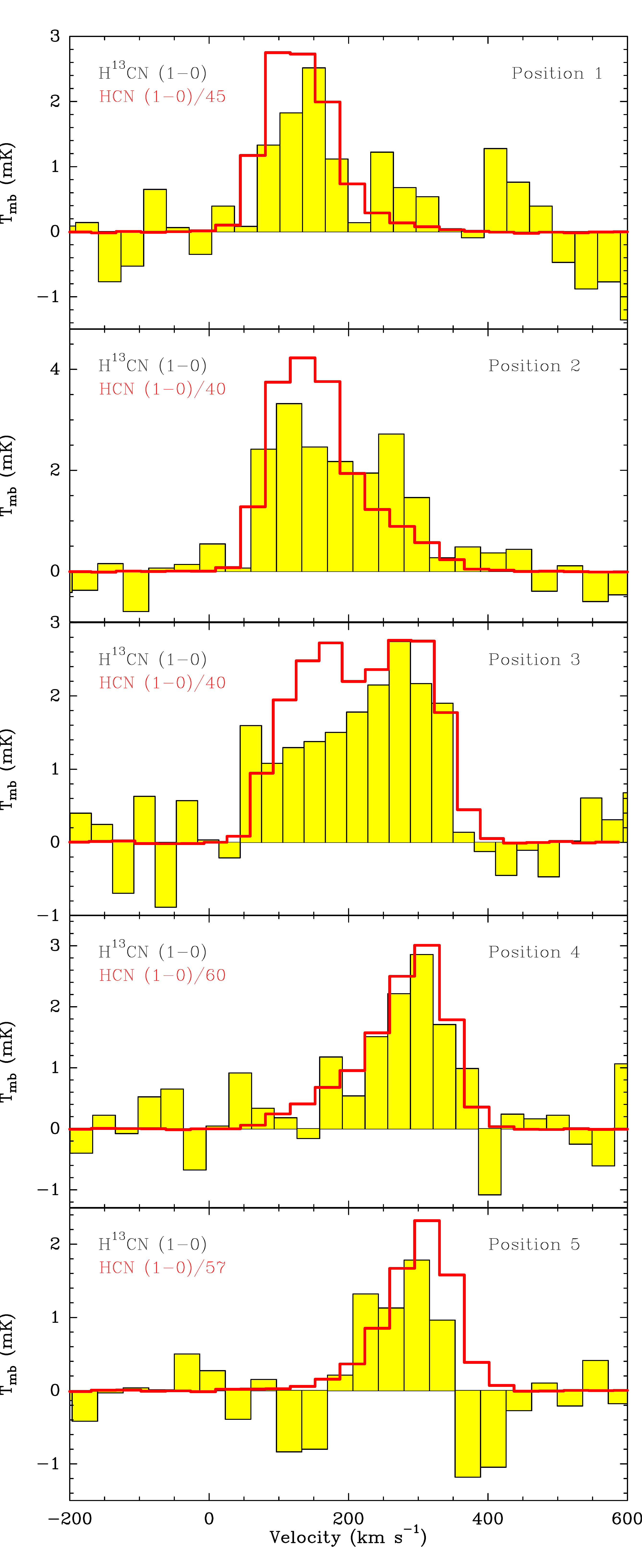}}
\subfigure {\includegraphics[scale=0.237 ]{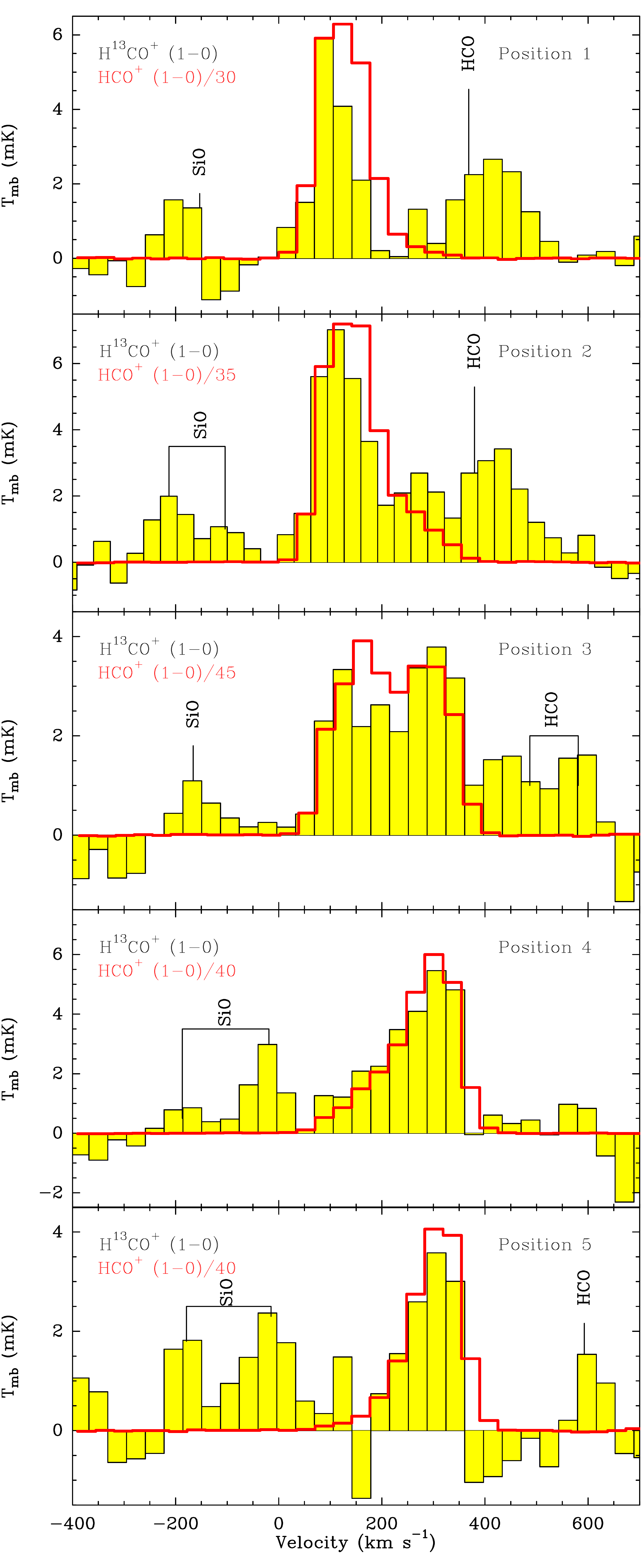}}

\vspace{2em}
\caption{
H$^{13}$CN $J$=1-0 (left) and H$^{13}$CO$^+$ $J$=1-0  (right) spectra (black histograms filled in
yellow)  measured at five positions along the major axis of M$\,$82. The HCN and
HCO$^+$ $J$=1-0 spectra are overlaid as red lines, after being normalised to
roughly the same peak intensities as the isotopic lines.  The velocity
resolution is $\sim$35 \kms\ for all spectra. The right panel has a slightly
larger velocity range to present the SiO and HCO spectra. }

\label{fig:H13CN_H13CO+}
\end{figure*}

\begin{figure*}
\centering
\subfigure {\includegraphics[scale=0.20]{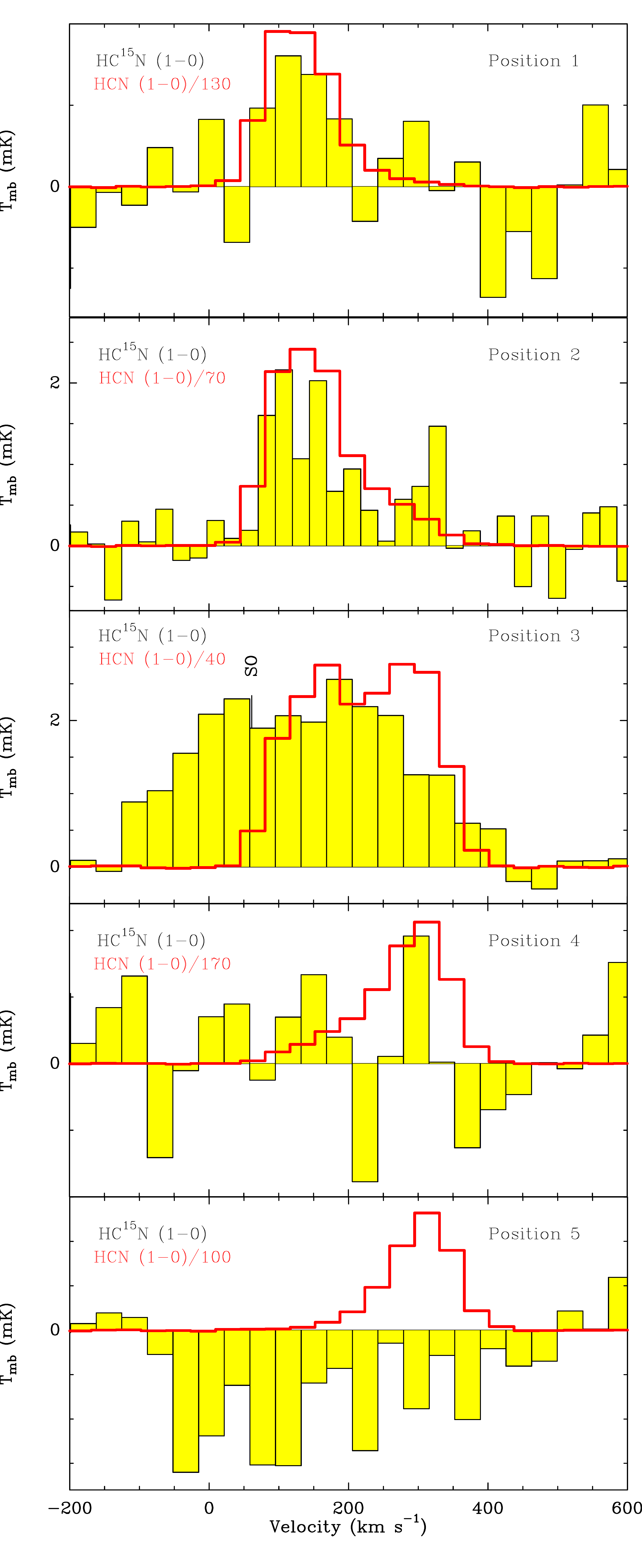}}
\subfigure {\includegraphics[scale=0.20]{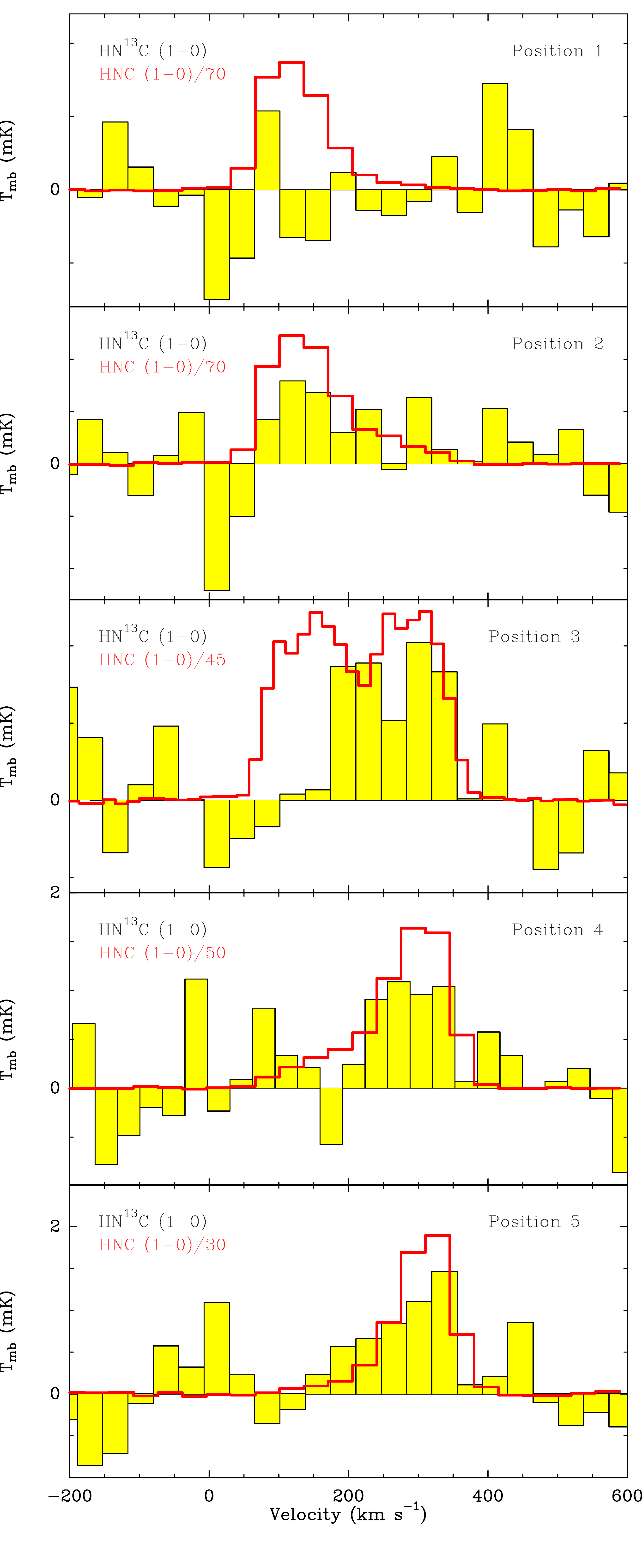}}
\subfigure {\includegraphics[scale=0.20]{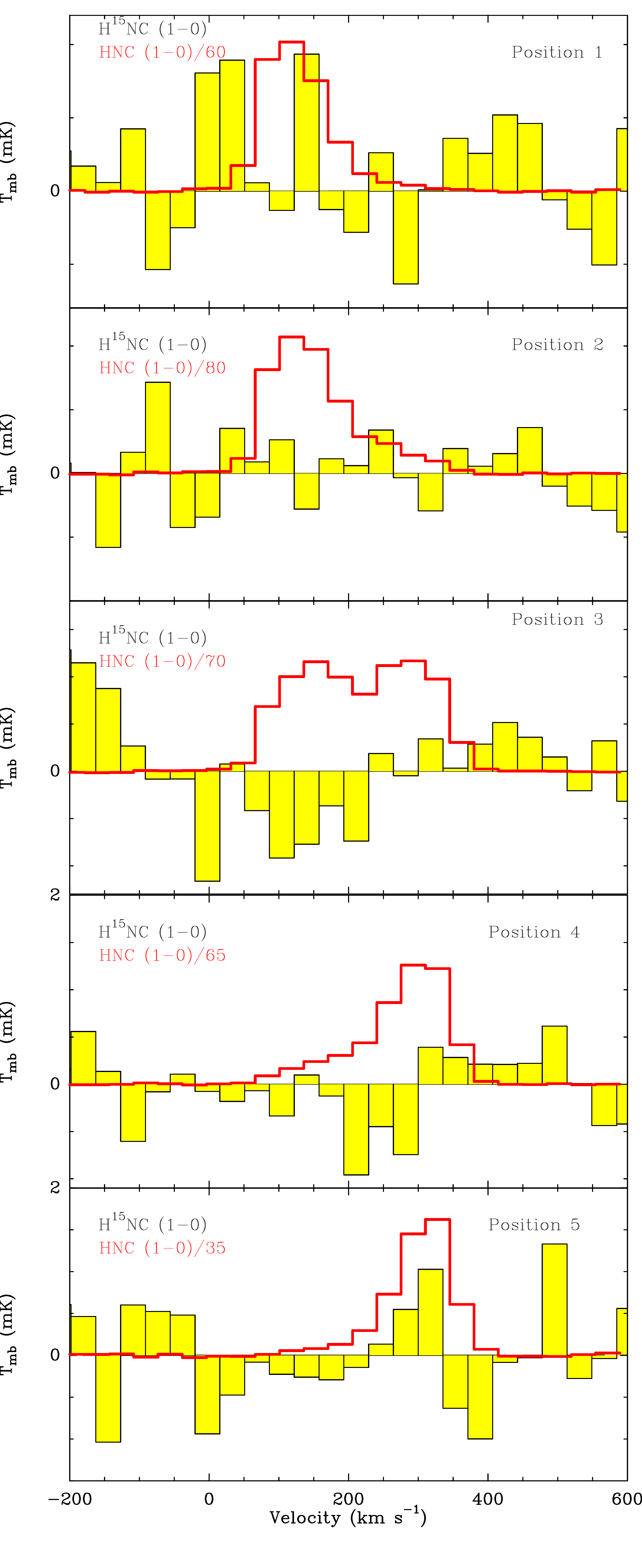}}
\vspace{2em}
\caption{HC$^{15}$N $J$=1-0  (left), HN$^{13}$C $J$=1-0  (middle)  and
H$^{15}$NC $J$=1-0 (right) spectra (black histograms filled in yellow)
measured at five positions along the major axis of M$\,$82. The HCN and HNC
$J$=1-0 spectra are overlaid as red lines, after being normalised to roughly
the same peak intensities as the isotopic lines. The velocity resolution is
$\sim$35 \kms\ for all spectra.  }
\label{fig:HC15N_HN13C_H15NC}
\end{figure*}

\begin{figure*}
\centering

\subfigure {\includegraphics[scale=0.20]{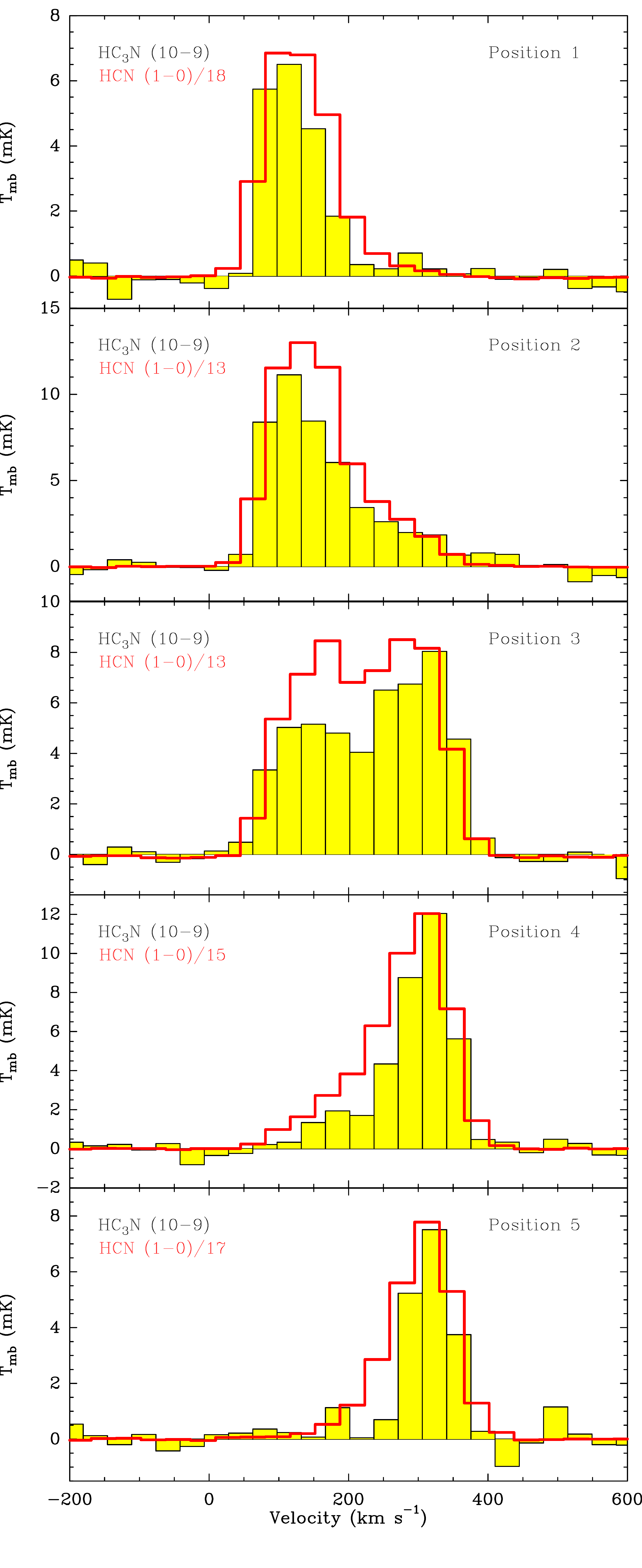}}
\subfigure {\includegraphics[scale=0.20]{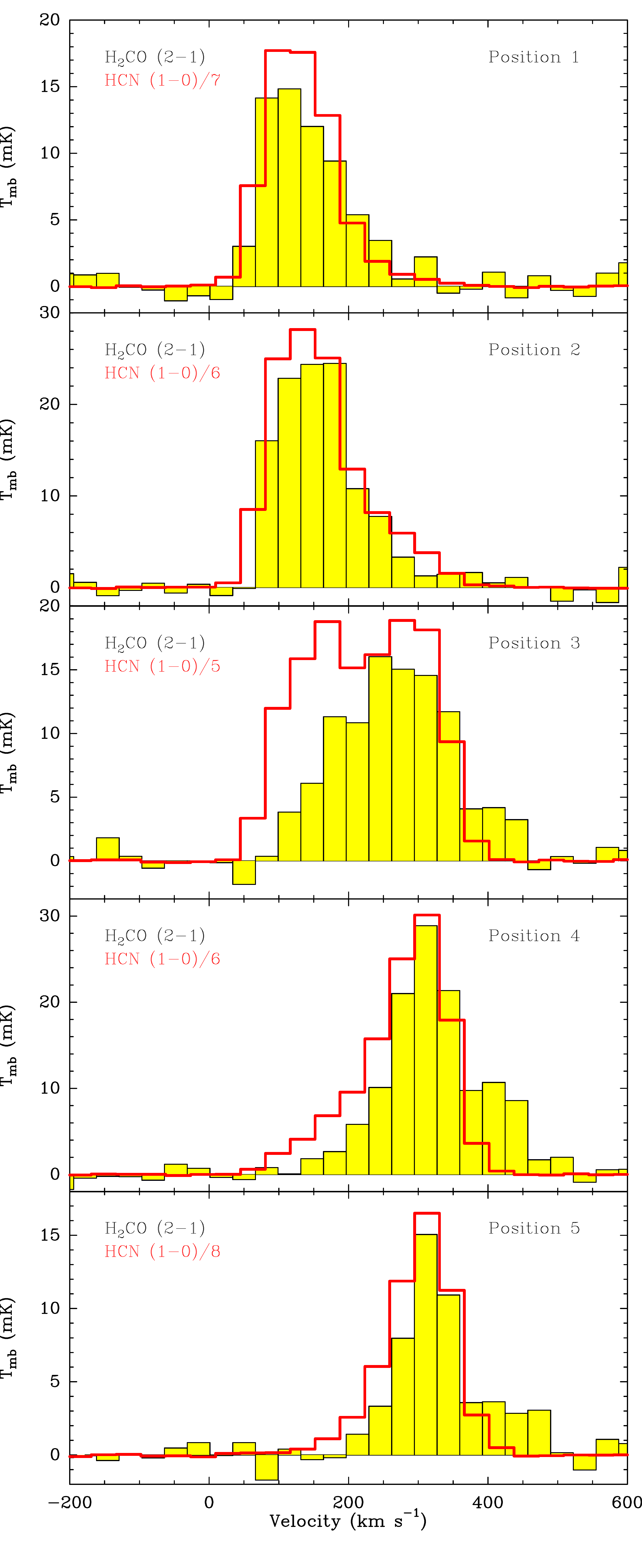}}
\subfigure {\includegraphics[scale=0.2]{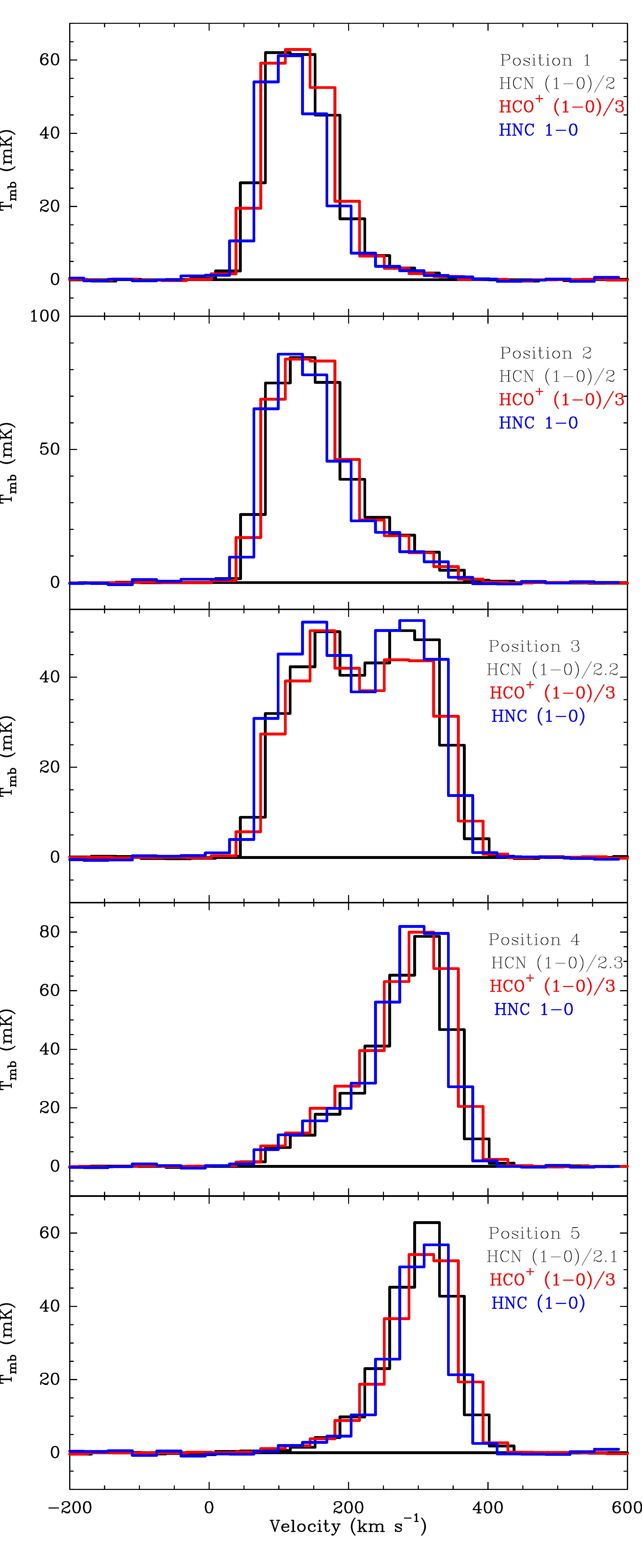}}
\vspace{2em}
\caption{
HC$_{3}N$ $J$=10-9 (Left), H$_2$CO 2-1 (middle) spectra, shown as black
histograms filled in yellow, measured at five positions along the major axis of
M$\,$82.  The HCN, HCO$^+$, and HNC $J$=1-0 spectra are overlaid with
black, red, and blue lines, respectively.  They are normalised to roughly the
same peak intensities. The velocity resolution is $\sim$35 \kms\ for all
spectra. 
}

\label{fig:HC3N_H2CO_Dense}
\end{figure*}

From the calibrated spectra data, we identify different transitions at each pointing 
position according to their rest frequency, 
which is listed in Table \ref{tab:all_position}.
Figures \ref{fig:H13CN_H13CO+}-\ref{fig:HC3N_H2CO_Dense} present spectra of
H$^{13}$CN $J$=1-0, HC$^{15}$N $J$=1-0, H$^{13}$CO$^+$ $J$=1-0, HN$^{13}$C
$J$=1-0, H$^{15}$NC $J$=1-0, and HC$_3$N $J$=10-9 covered by the 3-mm band,
while H$_2$CO $J$=2-1 is covered by the 2-mm band. The
intensities of all spectra are plotted on the $T\rm_{MB}$ scale.  For
comparison, spectra of the major isotopologues of HCN $J$=1-0, HCO$^+$ $J$=1-0
or HNC $J$=1-0 are overlaid.  

Most spectra, especially those from main isotopologues, at position-3
(hereafter P3; the central position) show two velocity components, which has
been also shown in \citet{Salas2014} and \citet{Li2020}. Only a single velocity component
can be identified from the line profiles at all four off-centre positions.
Among them, positions-1 and -2 (hereafter P1 \& P2; at the south-west side) are
dominated by the blue-shifted velocity component, while positions-4 and -5
(hereafter P4 \& P5; at the north-east side) are dominated by the red-shifted
velocity component, respectively.

\subsection{Detected lines}

As shown in \cite{Li2020}, the spectra at P3 can be fitted with two-component
Gaussian profiles.  However, the velocity difference between the two components
may not be constant among all five positions, and the outcome of the two-component Gaussian
fitting strongly depends on the data quality.  Therefore, we calculate the
velocity-integrated intensities of all molecular lines in two fixed velocity
ranges, i.e.,  0--200 \kms\ and 200--400 \kms, for the blue-shifted and the
red-shifted component, respectively. Table \ref{tab:all_position} presents the
velocity-integrated intensities at all positions. For P1 and P5, only the
major velocity component is considered, because of the weakness of the minor
component. Although some lines show emission line features on $\sim$ 2-$ \sigma$
levels, we only adopt 3-$\sigma$ at their upper limits. In the following, we
present properties of each isotopologue line.

\begin{itemize}
\item {\bf H$^{13}$CN $J$=1$\rightarrow$0, HN$^{13}$C $J$=1$\rightarrow$0, and H$^{15}$NC $J$=1$\rightarrow$0:} 

We present the line profile of H$^{13}$CN $J$=1-0 in the left panel of Figure \ref{fig:H13CN_H13CO+}.
H$^{13}$CN $J$=1$\rightarrow$0 is detected at all five positions. Two velocity
components are detected at three central positions, P2, P3, and P4. At P4, the
blue-shifted component only shows a weak feature at $\sim$ 2-$\sigma$ level.
Therefore, we present 3-$\sigma$ upper limits for it in Table
\ref{tab:all_position}.  HN$^{13}$C $J$=1-0 is detected in P3, P4, and P5, 
which is shown in the middle panel of Figure \ref{fig:HC15N_HN13C_H15NC}. The
north-east (NE) side shows stronger HN$^{13}$C emission than those at the
south-west (SW) side.  H$^{15}$NC $J$=1-0 is not detected at all positions, 
which is shown in the right panel of Figure \ref{fig:HC15N_HN13C_H15NC}. So,
we show 3-$\sigma$ upper limits of the velocity-integrated intensities in Table
\ref{tab:all_position}.

\item  {\bf HC$^{15}$N $J$=1$\rightarrow$0:}  

We present the line profile of HC$^{15}$N $J$=1-0 in the left panel of 
Figure \ref{fig:HC15N_HN13C_H15NC}.
HC$^{15}$N $J$=1$\rightarrow$0 is blended with SO $J$=2-1, with only a 200
\kms\ offset in velocity. This line blending effect is mostly severe at P3, where
SO $J$=2-1 peak is at the same level of HC$^{15}$N $J$=1$\rightarrow$0 (see
Figure \ref{fig:HC15N_HN13C_H15NC}). At all other four positions, however,
HC$^{15}$N $J$=1-0 emission seems not heavily polluted by the SO $J$=2-1
emission, possibly because the SO emission is only enhanced in the central
shocked region \citep{Pineau1993,Aladro2011b}. Therefore, we use the HCN line
profile at P3 as a template and fit the SO profile at the same position, then
remove the fitted SO contribution. We then use the residual to derive the
integrated intensity of HC$^{15}$N at P3. For P4 \&\, P5, we estimate
3-$\sigma$ upper limits for the velocity-integrated intensities.  Comparing
with the spatial variation of H$^{13}$NC $J$=1-0, HC$^{15}$N
$J$=1$\rightarrow$0 show a contrary trend along the major axis.


\item {\bf H$^{13}$CO$^+$ $J$=1$\rightarrow$0, HCO $J$=1$\rightarrow$0 and SiO $J$=2$\rightarrow$1:}

As shown in the right panel of Figure \ref{fig:H13CN_H13CO+}, H$^{13}$CO$^+$
$J$=1-0 is detected at all five positions. HCO $J$=1-0 and H$^{13}$CO$^+$
$J$=1-0 have a velocity offset of 294 \kms, therefore they show a weak blending
at P2 and P3. The HCO $J$=1-0 line is well detected in the SW side, while in
the NE side it was only marginally detected at P5.

SiO $J$=2-1 seems not blended with H$^{13}$CO$^+$ $J$=1-0. SiO $J$=2-1 is
stronger at the NE side than that at the SW side, where we only obtain
3-$\sigma$ upper limits at P1\&\,P2. This asymmetric distribution of SiO
indicates a fast shock at the NE side, possibly driven by outflow
\citep{Gusdorf2008,Gibb2007,Lopez2011}.

\item {\bf HC$_3$N $J$=10$\rightarrow$9:} 

We present the line profile of HC$_3$N $J$=10-9 in the left panel of 
Figure \ref{fig:HC3N_H2CO_Dense}.
Being optically-thin in most galactic environments \citep{Morris1976}, HC$_3$N
10-9 is detected at all positions. This line is stronger than those
isotopologue lines of other dense gas tracers at the same position. The line
profile of HC$_3$N is similar to that of HCN $J$=1$\rightarrow$0 at each
position. The velocity-integrated intensity of HC$_3$N $J$=10-9 at P-2 is
consistent with that reported by \cite{Aladro2015}.

\item {\bf H$_2$CO $J$=2(0,2)-1(0,1):} 
We present the line profiles of  H$_2$CO $J$=2(0,2)-1(0,1) in the middle panel of
Figure \ref{fig:HC3N_H2CO_Dense}. Most of the emission feature around 145.603
GHz seems from H$_2$CO $J$=2(0,2)-1(0,1), whose frequency fits the line profile
very well.  However, HC$_3$N $J$=16-15 might also contribute part of the flux to
the red-shift emission feature. 
       
\item {\bf HCN, HCO$^+$, and HNC $J$=1-0:} 
HCN, HCO$^+$ and HNC $J$=1-0 are all detected with high signal to noise level,
at all five positions. Their line profiles agree well with each other at the
same positions (see Figure \ref{fig:HC3N_H2CO_Dense}).  All these lines show
double-peak profiles at the central position, which is not always seen in those
weak isotopologue lines.

\end{itemize}
 
%
%
%

\begin{figure*}
\centering

\subfigure {\includegraphics[height=1.8in,width=3in]{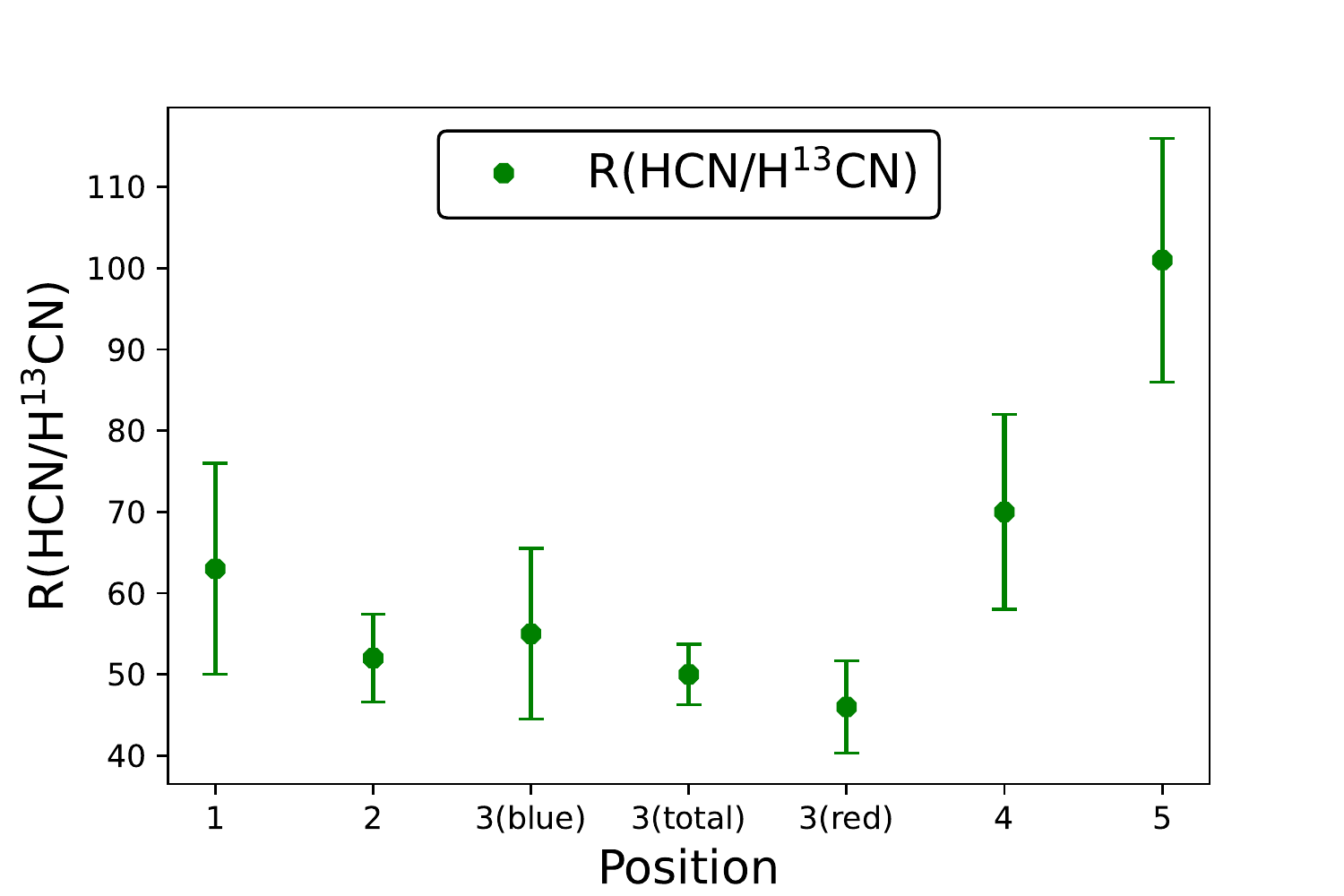}}
\subfigure {\includegraphics[height=1.8in,width=3in]{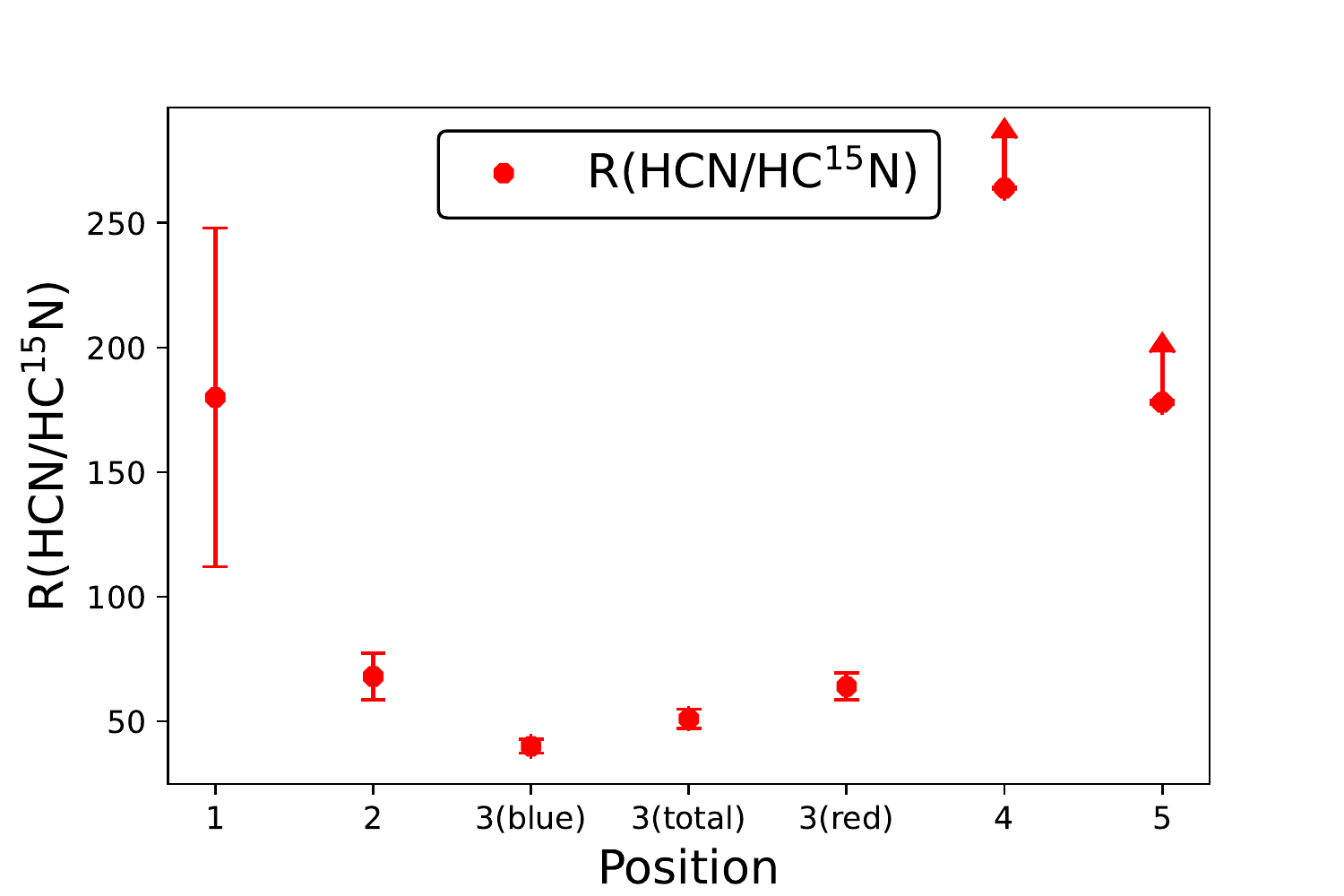}}
\subfigure {\includegraphics[height=1.8in,width=3in]{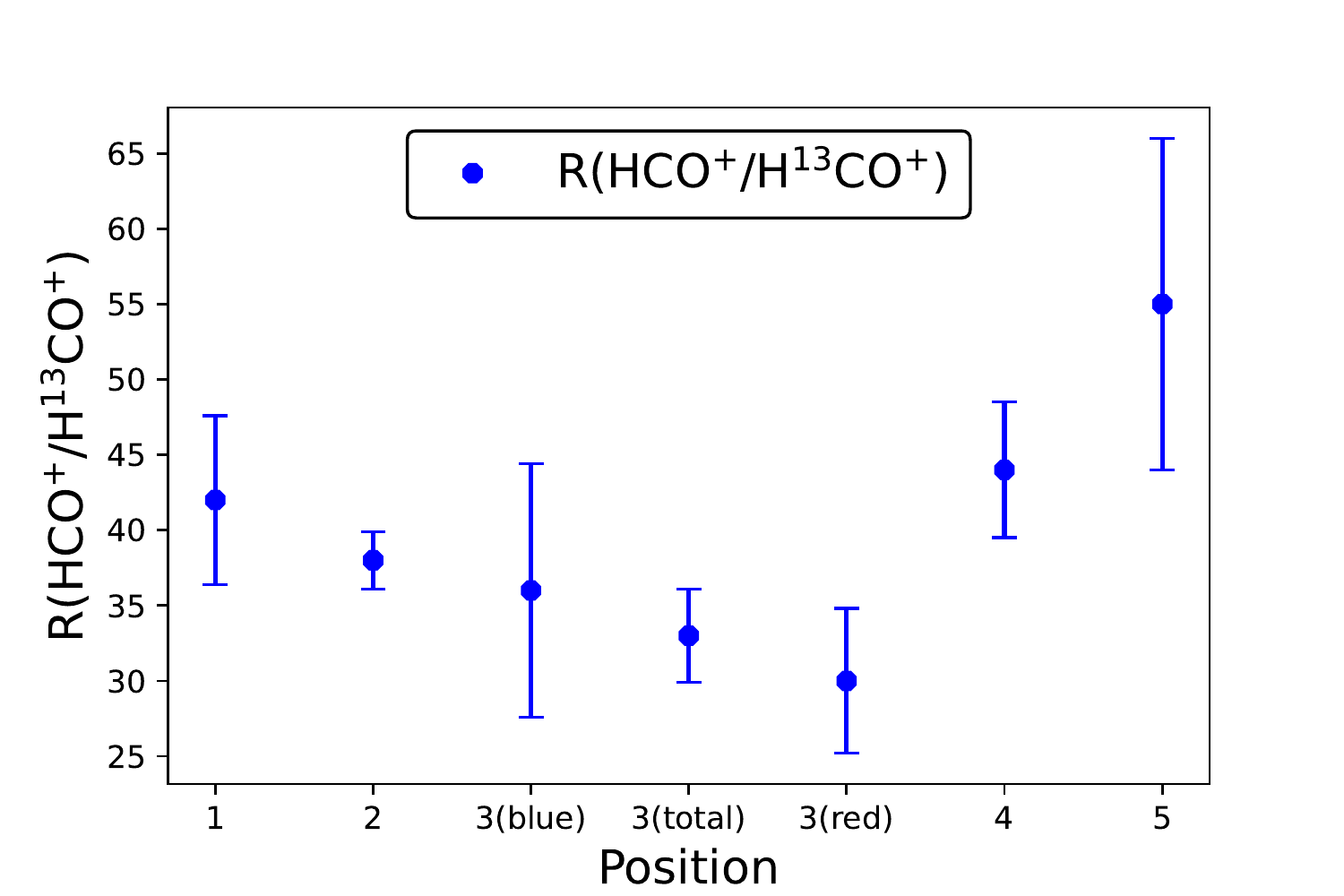}}
\subfigure {\includegraphics[height=1.8in,width=3in]{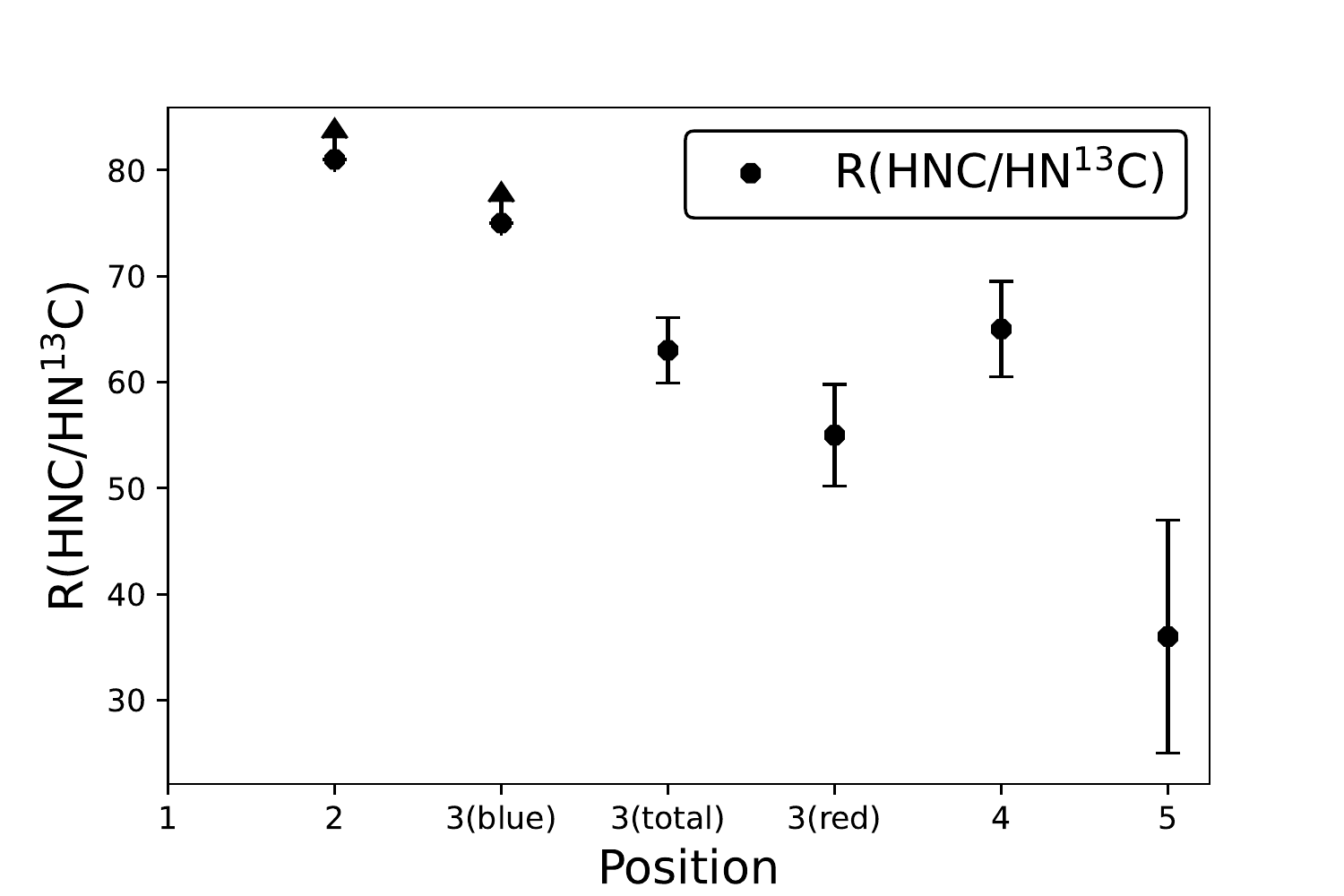}}

\vspace{2em}
\caption{The isotopic ratios  as a function of Positions. These isotopic ratios are $I$(HCN)/$I$(H$^{13}$CN), $I$(HCN)/$I$(HC$^{15}$N),$I$(HCO$^+$)/$I$(H$^{13}$CO$^+$), and $I$(HNC)/$I$(HN$^{13}$C), respectively}
\label{fig:Ratio_position}
\end{figure*}

\subsection{ Line Intensity Ratios}


With all measured velocity-integrated intensities, we obtained line intensity ratios between
the main and rare isotopologue lines of HCN, HCO$^{+}$, and HNC at each position, 
and we list results in Table \ref{tab:Isotopic_ratio}. At the central position (P-3), we
calculated ratios obtained from three velocity-integrated intensities, i.e.,
blue-shifted, red-shifted , and total velocity components of these lines. 

As shown in Table \ref{tab:Isotopic_ratio}, the ratios of $I$(H$^{13}$CO$^{+}$)/$I$(H$^{13}$CN) and
$I$(H$^{13}$CN/$I$(HC$_{3}$N) seem relatively constant among all positions,
with a variation $\sim$20-30\%. However, most other intensity ratios show a
significant spatial variation,  which can reach up to a factor of 4--5 (see Figure
\ref{fig:Ratio_position}).  The spatial variation of $I$(HCN)/$I$(H$^{13}$CN),
$I$(HCO$^{+}$)/$I$(H$^{13}$CO$^{+}$), and $I$(HCN)/$I$(HC$^{15}$N) ratios
follow a similar trend, with higher values at the NE side than those at the SW
side.  These ratios in the central region are lower than those on the disk.
These trends are consistent with \textbf{what has been} found in $I(^{12}$CO)/$I(^{13}$CO)
$J$=1-0 \citep{Kikumoto1998}.  However, the $I$(HNC)/$I$(HN$^{13}$C) intensity
ratios show an opposite trend along the major axis, compared with others.

\subsection{Optical depths of the main isotopologue lines}

Under the assumptions of local thermal equilibrium (LTE), uniform gas excitation,
constant isotopic abundance ratios, and neglecting radiative transfer by the
cosmic microwave background, we can derive optical depths of the main isotopic
lines. We adopt the same method as reported in \cite{Li2020},
 in which the line ratio $R$ and the optical depth for a specific isotopologue 
 line $\tau$, is related as$R$ =
(1-$e^{-\tau_{12}}$)/(1-$e^{-\tau_{13}}$)

Due to the large uncertainty of absolute abundance ratio of isotopes in external galaxies,
 we adopt various $^{12}$C/$^{13}$C and $^{14}$N/$^{15}$N
abundance ratios reported in the literature:  $^{12}$C/$^{13}$C = 89
\citep[the Solar system,][]{Clayton2004} and 200
\citep[Ultra-Infrared-luminous ULIRGs,][] {Romano2017}), and $^{14}$N/$^{15}$N
= 100  \citep[nearby starburst galaxies,][]{Chin1999}, 200 \citep[Galactic massive
star-forming regions,][]{Li2017}, and 290 \citep[the Solar neighbourhood at
$R_{\rm gc}\sim 7.9$ kpc,][]{Adande2012}. 
Table \ref{tab:depth_HCN} shows the derived optical depths of HCN, HCO$^+$, and HNC
$J$ = 1$\rightarrow$0.

The optical depths of HCN $J$=1-0 and HCO$^{+}$ $J$=1-0 only have a slight
variation along the major axis of M$\,$82, with higher optical depths in the
central region than those on the two sides of the disk. The observed
I(HCN)/I(H$^{13}$CN) $J$=1-0 line ratio could exceed the assumed
$^{12}$C/$^{13}$C abundance ratio occasionally, meaning that the actual $^{12}$C/$^{13}$C
ratio should be even higher. Therefore, the optical depths calculated with
$^{12}$C/$^{13}$C = 89 are the lower limits, making $\tau_{\rm HCN\, J=1-0}>1$
at most positions. This also applies to the $\tau_{\rm HCN\, J=1-0}$ calculated
with the assumptions of various $^{14}$N/$^{15}$N abundance ratios.  Due to
limited sensitivity, we can only obtain optical depths of HNC $J$=1-0 at three
positions.

\subsection{Constraints \textbf{on} $^{12}$C/$^{13}$C and $^{14}$N/$^{15}$N Abundance Ratios}

To obtain optical depths accurately, one should also consider the possible
radial gradient of $^{12}$C/$^{13}$C abundance ratio, which increases radially in our Milky Way.
 This is currently not possible due to the
lack of abundance measurements. However, our newly measured line ratios between
the isotopologues can still be used to set tight constraints on the abundance ratios. 
The observed HCN/H$^{13}$CN line ratio reaches 101$\pm$15, which is
higher than the $^{12}$C/$^{13}$C abundance ratio in the Solar neighbourhood
($\sim$ 89). But it is consistent with the values measured in the outer
Galactic disk regions \citep{Milam2005}. 

With the double isotope method \citep{Adande2012}, the $^{14}$N/$^{15}$N ratio
can be calculated using the line ratio of $I$(H$^{13}$CN)/$I$(HC$^{15}$N) and
the $^{12}$C/$^{13}$C abundance ratio,  by $^{14}$N/$^{15}$N =
$I$(H$^{13}$CN)/$I$(HC$^{15}$N) $\times$  $^{12}$C/$^{13}$C.  Assuming
$^{12}$C/$^{13}$C=89 in at SW side and at the center, average values of
$^{14}$N/$^{15}$N ratios are 186 $\pm$ 113 and 91$\pm$11, respectively. On the
other hand, at the NE side, adopting $^{12}$C/$^{13}$C=89, the lower limits of
$^{14}$N/$^{15}$N ratios range from 156 to 334.

 \begin{table*}
\centering

      \caption{Intensity ratios of the isotopologues of HCN, HNC and HCO$^+$}   
      \label{tab:Isotopic_ratio}
      \begin{minipage}{160mm}
        \begin{tabular}{llllllllllllllll}
             \hline
            \hline
             
             
 {Position}               &{$\frac{I( \rm HCN)}{I( \rm H^{13}CN)}$} & {$ \frac{I( \rm HCN)}{ I( \rm HC^{15}N)}$} & {$ \frac{ I( \rm HCO^{+})}{ I( \rm H^{13}CO^{+})}$} & {$\frac{ I( \rm HNC)}{ I( \rm HN^{13}C)}$} & {$ \frac{ I( \rm H^{13}CN)}{ I( \rm HC^{15}N)}$} & {$ \frac{ I( \rm H^{13}CO^+)}{ I( \rm H^{13}CN)}$} & {$ \frac{ I( \rm H^{13}CN)}{ I( \rm HC_{3}N)}$} \\
\hline
                        &$J$=1-0                                  & 1-0                                        & 1-0                                                 & 1-0                                        & 1-0                                              & 1-0                                                & 1-0/10-9\\
\hline
1                       &63 $\pm$ 14                          & 180$\pm$   68                          & 42 $\pm$   6                                    & $>66  $                               & 2.89$\pm$1.25                             & 2.30 $\pm$   1.25                               & 0.36   $\pm$   0.08    \\
2                       &52 $\pm$ 5                           & 68 $\pm$   9                           & 38 $\pm$   2                                    & $>81  $                               & 1.31$\pm$0.23                             & 2.10 $\pm$   0.23                               & 0.32   $\pm$   0.04    \\
3$_{~\rm a} $           &55 $\pm$ 11                          & 40 $\pm$   3                           & 36 $\pm$   3                                    & $>75  $                               & 0.72$\pm$0.15                             & 2.38 $\pm$   0.15                               & 0.32   $\pm$   0.06    \\
3$_{~\rm b} $           &46 $\pm$ 6                           & 64 $\pm$   5                           & 32 $\pm$   2                                    & 55$\pm$13                             & 1.37$\pm$0.20                             & 2.15 $\pm$   0.20                               & 0.31   $\pm$   0.04    \\
3$_{~\rm total} $       &51 $\pm$ 6                           & 52 $\pm$   3                           & 34 $\pm$   2                                    & $>65$                                 & 1.02$\pm$0.13                             & 2.24 $\pm$   0.13                               & 0.32   $\pm$   0.02    \\
4                       &70 $\pm$ 12                          & $>264  $                               & 44 $\pm$   5                                    & 65$\pm$22                             & $>3.75 $                                  & 2.26 $\pm$   0.44                               & 0.26   $\pm$   0.04    \\
5                       &101$\pm$ 15                          & $>178  $                               & 55 $\pm$   11                                   & 36$\pm$11                            & $>1.75 $                                 & 2.57 $\pm$   0.62                               & 0.23   $\pm$   0.04    \\
\hline
 
\end{tabular}\\
Note. To keep uniformity, we only calculate the intensity ratio of one velocity
component. \\
$^{~\rm a} $ The ratios of the blue-shifted emission are estimated for
positions-1 and -2.\\
$^{~\rm b} $For positions-3, -4, and -5, the red-shifted emission are used to
calculate the ratios.\\
\end{minipage}
\end{table*}

\begin{table*}
\footnotesize
\centering
    \tabcolsep 1mm
    
      \caption{Optical depths for HCN, HNC, and HCO$^+$}  
      \label{tab:depth_HCN}
      \begin{minipage}[c]{160mm}
                \begin{tabular}{llllllllllllll}
             \hline
            \hline

{Position}  &    \multicolumn{2}{c}{${\tau}$(HCN)$^{~\rm a}$ (1-0) }  & &  \multicolumn{2}{c}{ ${\tau}$(HNC)$^{~\rm b}$ (1-0)}  &  & \multicolumn{2}{c}{$\tau$(HCO$^+$)$^{~\rm c}$ (1-0) }    &    & \multicolumn{3}{c}{${\tau}$(HCN)$^{~\rm d}$ (1-0)}  \\                                             
\cline{2-3} \cline{5-6} \cline{8-9} \cline{11-13} 
&     89$^{~\rm e}$  &  200$^{~\rm e}$                & & 89 &  200   & &   89 &  200         &  &   100$^{~\rm f}$ & 200$^{~\rm f}$ &  290$^{~\rm f}$ \\    
\hline                                                                                                                              
1                 & 0.75$\pm$0.52   & 3.05$\pm$0.82   & & $<$0.64        & $<$2.88           &  & 1.78$\pm$0.38        & 4.78$\pm$0.68       &  & $-$           & 0.22$\pm$0.01        & 1.05$\pm$0.13    \\           
2                 & 1.21$\pm$0.27   & 3.79$\pm$0.44   & & $<$0.19        & $<$2.21           &  & 2.07$\pm$0.15        & 5.31$\pm$0.28       &  & 0.84$\pm$0.39 & 2.77$\pm$2.18        & 4.23$\pm$1.40    \\                  
3$_{~\rm a} $     & 1.07$\pm$0.48   & 3.57$\pm$0.79   & & $<$0.36        & $<$2.45           &  & 2.24$\pm$0.26        & 5.61$\pm$0.75       &  & 2.27$\pm$1.40 & 5.03$\pm$3.10        & 7.34$\pm$2.28    \\                                  
3$_{~\rm b} $     & 1.52$\pm$0.34   & 4.34$\pm$0.58   & & 1.07$\pm$0.02  & 3.57$\pm$0.03     &  & 2.62$\pm$0.18        & 6.34$\pm$0.56       &  & 0.98$\pm$0.45 & 2.99$\pm$2.27        & 4.51$\pm$1.50    \\                
3$_{~\rm total} $ & 1.26$\pm$0.19   & 3.88$\pm$0.47   & & $<$0.75        & $<$3.05           &  & 2.41$\pm$0.15        & 5.95$\pm$0.31       &  & 1.51$\pm$1.21 & 3.80$\pm$2.61        & 5.61$\pm$1.82    \\                
4                 & 0.51$\pm$0.38   & 2.68$\pm$0.58   & & 0.67$\pm$0.81  & 2.93$\pm$1.3      &  & 1.65$\pm$0.28        & 4.55$\pm$0.5        &  & $-$           & $-$                  & $<$0.19         \\                   
5                 & $-$             & 1.58$\pm$0.40   & & 2.24$\pm$0.03  & 5.61$\pm$0.05     &  & 1.07$\pm$0.49        & 3.57$\pm$0.8        &  & $-$           & $<$0.24              & $<$1.07         \\                   
\hline                                                                                                                                                             
mean              & 1.11 $\pm$ 1.51 & 3.64 $\pm$ 2.64 & & $<0.75$        & $<3.05$           &  & 1.78 $\pm$ 0.98      & 4.78 $\pm$1.79       &  & $<1.11$       & $<0.9$               & $<2.12$\\                      
whole             & 1.16 $\pm$ 0.15 & 3.72 $\pm$ 0.24 & & $<0.64$        & $<2.88 $          &  & 1.92 $\pm$ 0.11      & 4.03 $\pm$0.21       &  & $<1.87$       & $<1.9$               & $<3.05$\\                         
\hline

\end{tabular}\\
\begin{list}{}{}
\item[${\mathrm{a}}$] The optical depth of HCN is calculated by the ratio of HCN/H$^{13}$CN.
\item[${\mathrm{a}}$] The optical depth of HNC $J$=1-0 is calculated by the ratio of HNC/HN$^{13}$C $J$=1-0.
\item[${\mathrm{b}}$] The optical depth of HCO$^+$ 1-0 is calculated by the ratio of HCO$^+$/H$^{13}$CO$^+$ 1-0.
\item[${\mathrm{d}}$] The optical depth of HCN is calculated by the ratio of HCN/HC$^{15}$N.
\item[${\mathrm{e}}$] Abundance ratios of $^{12}$C/$^{13}$C:  89 (Solar system \citep{Clayton2004} ) and 200 (ULIRGs \citep{Romano2017}).
\item[${\mathrm{f}}$] Abundance ratios of $^{14}$N/$^{15}$N: 100 (nearby starburst galaxies  \citep{Chin1999}), 200 ( Galactic massive star-forming regions \citep{Li2017}) and 290 (local interstellar medium \citep{Adande2012}).

\end{list}

 \end{minipage}
 \end{table*}

\section{Discussion} \label{sec:discussion}

\subsection{Comparison with data in the literature}

Table \ref{tab:Isotopic_ratio_other_galaxies} summarizes literature measurements
of $I$(HCN)/ $I$(H$^{13}$CN) $J$=1-0, $I$(HCO$^+$)/ $I$(H$^{13}$CO$^+$)
$J$=1-0, and $I$(HNC)/ $I$(HN$^{13}$C) $J$=1-0 from nearby galaxies, including
starburst galaxies, ULIRGs and active galactic nucleus (AGN)-dominated
galaxies. All these ratios in M$\,$82 are higher than those found in the
literature, except for M~83.

Previous observations of dense gas tracers have shown that they are mostly
optically thick in both Galactic giant molecular clouds (GMCs) and external
galaxies \citep{Wang2014,Meier2015,Jimenez2017,Li2017,Li2020}.  If we assume
that the abundance ratio of $^{12}$C/$^{13}$C is 40, which is obtained as the
average condition of nearby galaxies \citep{Henkel1994,Henkel2010}, the HCN,
HCO$^+$ and HNC lines of M~82 would be optically thin.

At all five positions \textbf{in} M$\,$82, ratios of $I$(H$^{13}$CN)/$I$(H$^{13}$CO$^+$)
$J$=1-0 are lower than those of $I$(HCN)/$I$(HCO$^+$) $J$=1-0. Similar results
for the $J$=2-1 transition were also reported by \cite{Aladro2011b} at P2, where the $I$(H$^{13}$CN)/$I$(H$^{13}$CO$^+$) $J$=2-1 ratio is
lower than that of $I$(HCN)/$I$(HCO$^+$) $J$=2-1. This indicates that HCN lines
should have lower optical depths than those of HCO$^+$ lines.

\subsection{Optical Depths }

Given the high intensity ratios of $I$(HCN)/$I$(H$^{13}$CN) $J$=1-0 in a range from 51$\pm$6
to 101$\pm$15 at all five positions, a $^{12}$C/$^{13}$C abundance ratio of 40
\citep{Henkel1998} can not be valid. \cite{Kikumoto1998} and \cite{Tan2011}
also found similar results, and they adopted a $^{12}$C/$^{13}$C abundance
ratio of 60.  No matter which $^{12}$C/$^{13}$C abundance ratio is adopted,
$^{12}$C-bearing dense gas tracer lines need to have at least moderate optical
depths.

The optical depths of dense gas tracers may be decreased by the feedback of
supernova explosions in M~82, which has been found to in
previous studies \citep{Allen1998, Mattila2001}.  Such supernova feedback
strongly affect the molecular gas environments, which can be seen from strong
SiO emission (see Figure \ref{fig:H13CN_H13CO+}), high H$_2$ 1-0 $S$(1)/Br$\gamma$ ratio, and high ratio of
$I$($^{12}$CO)/$I$($^{13}$CO) \citep{Lester1990, Mouri1989}. Therefore, the
$^{12}$C-bearing molecular lines would be expected to be more optically thin,
because the shock conditions could broaden the lines, which have an increased
escape probability in radiative transfer. Such effects have been seen in
Galactic supernova remanent, which are interacting with molecular clouds
\citep[e.g.,][]{Zhang2010},


To mimic spatially unresolved galaxies, we further derive the
weighted mean optical depths of dense gas tracers in M~82, by averaging optical
depths among all positions, weighted with their line fluxes (see Table
\ref{tab:depth_HCN} marked as ``mean'').  The unweighted mean optical depth,
which was obtained directly from the total flux ratios, is also listed in the
same tables (marked as ``whole''). Both ``mean'' and ``whole'' optical depths
agree well with those obtained on the disk, within the error bars. Thus, for
galaxies without spatially information, optical depths of dense gas tracers
obtained from a whole galaxy could generally represent average conditions on
the disk.

\subsection{ The \Ctw/\Cth\ abundance ratio }
\label{sec:12C_13C_Interprete}




The two isotopes of Carbon, \Ctw\ and \Cth, have different mechanisms of nucleosynthesis.
The main isotope, \Ctw, could be partly produced by the triple-$\alpha$ reaction in massive stars 
\citep[$>$ 8 M$_\odot$,][]{Wilson1992, Nomoto2013}, and partly by lower mass stars. 
Therefore, \Ctw\ can increase quickly  after the starburst starts, due to the short 
lifetime of massive stars. Most of \Cth, on the other hand, is formed in low- and 
intermediate-mass stars ($<$ 8 M$_\odot$), which means that most \Cth\ should be released 
to the ISM on much longer timescales than that of \Ctw\ \citep{Wilson1992,Hughes2008}.   

As a result, the \Ctw/ \Cth\ abundance ratio could represent the star-formation 
history \citep{Wilson1994, Henkel2010} ---  high ratios for long-term starbursts (a few hundreds of Myrs) 
and low ratios for young starbursts (tens of Myrs), before the low-mass stars start to release \Cth. 

Given the relatively short starburst timescales of M$\,$82 \citep[$\sim 5 \times 10^7$yr,][]{Konstantopoulos2009}, 
the majority of the newly born low-mass stars are still alive. So, the \Ctw/\Cth\ ratio can be enhanced in the central
regions, where starburst is more intensive compared to that in the outer disk. Therefore, the current starbursts generate the \Ctw\ / \Cth\ 
decreasing trend from center to outskirt along the major axis of M$\,$82.

However, the observed \Ctw/ \Cth\ ratios are not only contributed from the current starbursts contribution, but also produced from all the past starburst activities in history.
Besides,
Galactic chemical evolution models predict that the \Ctw/ \Cth\ ratio would vary by 
a factor of two, even if a strong starburst produce half of the stellar mass in a secular evolving 
galaxy \citep{Romano2017}. Even if the undergoing starburst could slightly enhance \Ctw/ \Cth\ abundance ratio in a short timescale in the central region during the starburst, it would not change the increasing \Ctw/ \Cth\  gradient in M~82, similar to the Milky Way.


\begin{table*}
\centering

      \caption{Integrated intensity ratios from the literature} 
      \label{tab:Isotopic_ratio_other_galaxies}
      \begin{minipage}{160mm}
      \begin{tabular}{llllllllll}

             \hline
            \hline
             
             {Galaxy}   & {$ \frac{\rm HCN}{\rm H^{13}CN}$} & {$ \frac{\rm HCO^{+}}{\rm H^{13}CO^{+}}$} & {$\frac{\rm HNC}{\rm HN^{13}C}$} & Type    & Reference \\
                        & 1-0                               & 1-0                                       & 1-0  \\

             \hline   
NGC~3079                & 10$\pm$ 5                       & 7$\pm$2                                   & $>$28                          & SB/AGN  & 1\\
Mrk~231                 & 16$\pm$5                        & 12$\pm$5                                    & $>$7                             & SB      & 1\\
NGC~4418                & 8$\pm$3                           & $-$                                       & $-$                              & SB/AGN  & 2\\
NGC~1068                & 16$\pm$1                          & 20$\pm$1                                  & 38$\pm$6                         & AGN     & 3\\
NGC~3351                & 21$\pm$3                          & $>$17                                     & $-$                              & SB(r)b  & 4 \\
NGC~3627                & 7$\pm$1                           & $>$12                                     & $-$                              & SB/AGN  & 4 \\
NGC~253                 & 17$\pm$1                          & 24$\pm$2                                  & $-$                              & SB      & 4\\
M~83                    & 41$\pm$7                        & 44 $\pm$13                                & $-$                              & SAB(s)c & 5 \\
NGC~5194                & 27  $\pm$18                       & 34$\pm$29                                 & $>$16                            & SB/AGN  & 6\\
M$\,$82                    &46$\pm$6$-$101$\pm$15                &32$\pm$2$-$55$\pm$11                          &36$\pm$11$-$65$\pm$22              &SB       &this work\\
\hline

        \end{tabular}\\

{References: (1)\cite{Li2020} (2) \cite{Costagliola2011} (3) \cite{Wang2014} (4)\cite{Jimenez2017} (5)\cite{Aladro2015} (6)\cite{Watanabe2014} (7)\cite{Muller2011}.}

 \end{minipage}
 \end{table*}

\subsection{ The $^{14}$N/$^{15}$N abundance ratio}

\begin{figure*}
\centering
\includegraphics[scale=0.5]{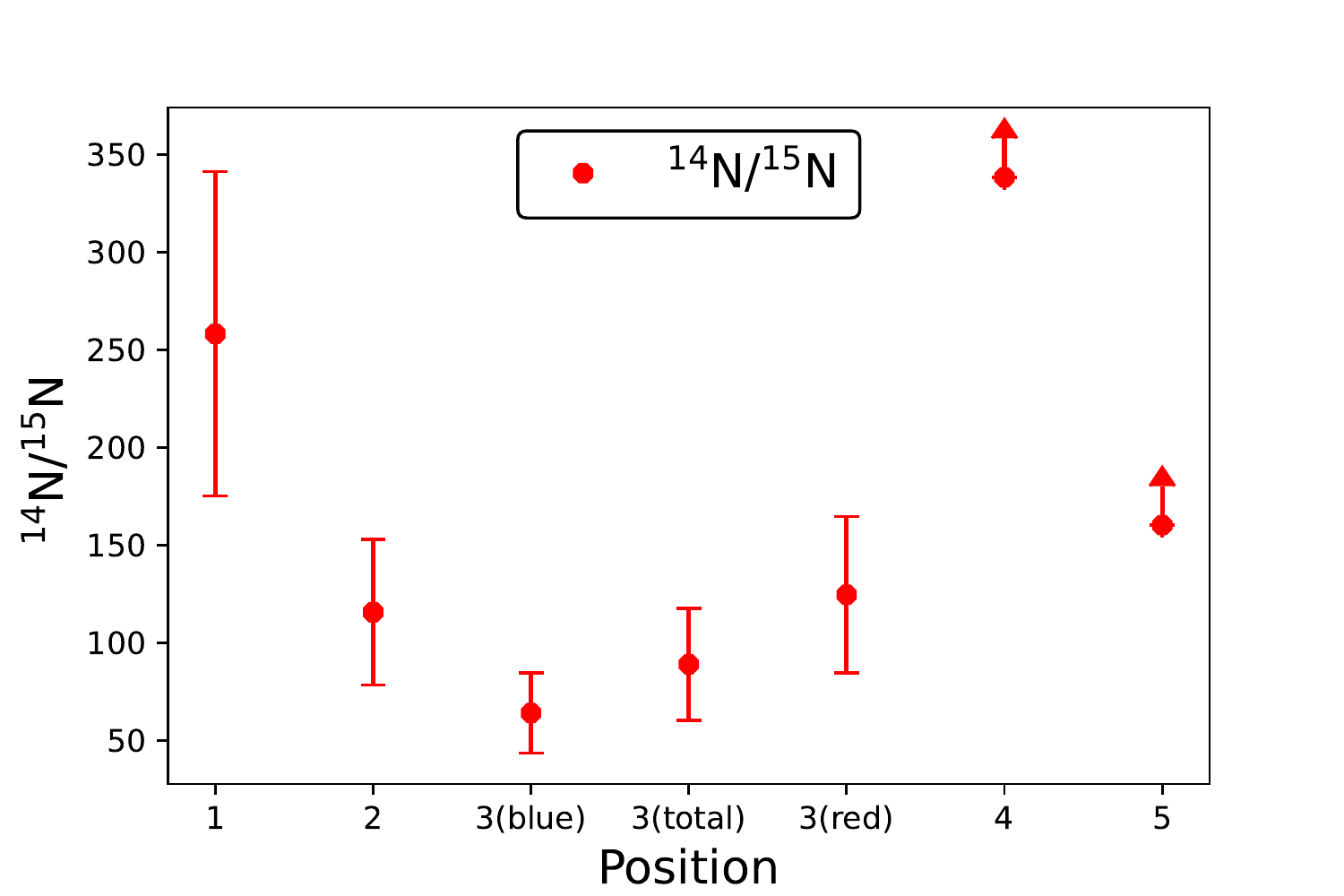}
\caption{The abundance ratio of $^{14}$N/$^{15}$N as a function of positions.}

\label{fig:14N_15N}
\end{figure*}

In external galaxies, the $^{14}$N/$^{15}$N abundance ratios have been measured
in a range from $\sim$ 100 to $\sim$ 450
\citep{Henkel1998,Henkel2018,Wang2014,Wang2016,Adande2012}. This ratio measured
in two starbursts,  NGC~4945 \citep[$\sim$200-- 400,][]{Henkel2018} and ULIRG
Arp~220 \citep[$440^{+140}_{-82}$,][]{Wang2014}, are higher than those found in
M~82 ($\sim 80 --250$, see Figure \ref{fig:14N_15N}).  In our Milky Way, the
$^{14}$N/$^{15}$N ratios distribute in a large range
\citep[100--600,][]{Li2017}, with a possible positive gradient from the center
to out disk \citep{Chen2021}. 



Figure \ref{fig:14N_15N} shows estimated $^{14}$N/$^{15}$N ratios as a function
 of galactocentric distance. If we adopt an assumption of constant $^{12}$C/$^{13}$C abundance
ratio of 89, $^{14}$N/$^{15}$N abundance ratios tend to increase with
galactocentric distance, which is consistent with the $^{14}$N/$^{15}$N
gradient found in the Milky Way
\citep{Dahmen1995,Adande2012,Romano2003,Romano2017,Chen2021}.

On the other hand, the $\rm ^{12}C/^{13}C$ abundance ratio shows an increasing
gradient from the center to the outer disk in the Milky Way
\citep{Wilson1994,Savage2002,Milam2005,YtYan2019}. If we adopt a similar
positive $^{12}$C/$^{13}$C abundance gradient to M~82, which is consistent with the prediction in 
Section \ref{sec:12C_13C_Interprete}, the $^{14}$N/$^{15}$N
abundance ratio would show an even stronger gradient, with lower values in the
center and higher ratios on the disk.

\subsection{Astrochemical effects}

Selective photon dissociation prefers to destroy $^{13}C$- and $^{15}C$-bearing
molecules, which would increase $\rm ^{12}C/^{13}C$ and $\rm ^{14}N/^{15}N$
abundance ratios in high UV fields \citep{Wilson1992,Savage2002}.  However, the
inner region of M$\,$82 has lower $I$(HCO$^+$)/$I$(H$^{13}$CO$^+$)  and
$I$(HCN)/$I$(HC$^{15}$N) line values, indicating that photo-dissociation does not
play a key role. 

Isotope fractionation, which could enhance H$^{13}$CO$^+$/HCO$^+$ and
HC$^{15}$N/HCN ratios, is only effective at a very low temperature
\citep{Smith1980,Woods2009,Rollig2013,Loison2019}. In M~82, the temperature is
relatively high, especially in the center. However, both
$I$(HCO$^+$)/$I$(H$^{13}$CO$^+$)  and $I$(HCN)/$I$(HC$^{15}$N) ratios increases
in the off-center regions, meaning that fractionation effect does not play a
key role as well.


\section{Summary}
\label{sec:summary}

In this paper we present results from IRAM 30-m telescope observations along the major axis of M$\,$82 in the 2-mm band and 3-mm band. Four positions are selected to measure the isotopic lines of HCN, HCO$^+$, and HNC. The spatial distribution of these optically thin dense gas tracers are obtained, including H$^{13}$CN, H$^{13}$CO$^+$, HC$^{15}$N, HN$^{13}$C, H$^{15}$NC 1-0. A few other species of SiO  $J$=2-1, HCO  $J$=1-0, H$_2$CO  $J$=2-1, and HC$_3$N  $J$=10-9 are also detected. We obtain the following results:

(1) H$^{13}$CN  $J$=1-0, H$^{13}$CO$^+$  $J$=1-0, HC$_3$N  $J$=10-9  and H$_2$CO  $J$=2-1 are detected in all four positions along the major axis of M$\,$82. Spectral lines of the transition HC$^{15}$N $J$=1-0 are not detected at the NE side and HN$^{13}$C $J$=1-0 emissions are not detected at the SW side. We did not obtain any detection of H$^{15}$NC $J$=1-0 at all positions
For the tentative detection of the isotopic lines, the 3-$\sigma$ upper limits are presented.



(2)For spectral line intensity ratios, $I$(HCN)/$I$(H$^{13}$CN) $J$=1-0, $I$(HCO$^+$)/$I$(H$^{13}$CO$^+$) $J$=1-0 and $I$(HCN)/$I$(HC$^{15}$N) $J$=1-0 show a large spatial variation along the major axis of M$\,$82, which are higher at the NE side than those at the SW side and the value in the central region is lower than that on the disk. However, the $I$(HNC)/$I$(HN$^{13}$C) ratio seems to show an opposite trend along the major axis.

(3) 
The optical depths of HCN $J$=1-0 and HCO$^{+}$ $J$=1-0 only have a slight
variation along the major axis of M$\,$82, with higher optical depth in the
central region than those on the two sides of the disk. Due to limited
sensitivity, we can only obtain optical depths of HNC $J$=1-0 at three
positions, thus could not summarise any trend for its optical depth variation.


(4) Our measured line ratios between the isotopologues set a lower limit for the abundance ratios of $^{12}$C/$^{13}$C. Using the double method and $I$(H$^{13}$CN/$I$(HC$^{15}$N) ratio,  the derived $^{14}$N/$^{15}$N abundance ratios have an increasing gradient from the center to the outer disk.


\section*{Acknowledgements}

This work is supported by the National Natural Science Foundation of China
grant (12041305, 12173067 and 121030243), and the fellowship of China Postdoctoral
Science Foundation 2021M691531.  We would like to thank P. Salas to provide
their data for Figure \ref{fig:M82_cor}.  We are grateful to the staff of IRAM
30-m telescope for their kind help and support during our observations. This
study is based on observations carried out under project number 186-18 (PI: Feng Gao) with the
IRAM 30-m telescope. IRAM is supported by INSU/CNRS (France), MPG (Germany),
and IGN (Spain).  We acknowledge the Program for Innovative
Talents, Entrepreneur in Jiangsu. We acknowledge the science
research grants from the China Manned Space Project with NO.CMS-CSST-2021-A08.

\appendix

\bibliographystyle{mnras}
\bibliography{M82}

\appendix

%
%
%
%
%

\section{Other lines:SiO $J$=2-1, HCO $J$=1-0, H$_2$CO $J$=2-1, and HC$_3$N $J$=10-9 } 
\label {sec:Appendix A}

Apart from the isotopic lines, we also detected a few unique molecular lines,
such as SiO $J$ = 2$\rightarrow$1, HCO $J$ = 1$\rightarrow$0, H$_2$CO $J$ =
$\rightarrow$1, and HC$_3$N $J$ = 10$\rightarrow$9. These lines offer an
excellent opportunity to reveal physical conditions of the ISM of M$\,$82.

\subsection{SiO $J$=2-1}

SiO, as the shock tracer \citep{Usero2006}, is detected in both the central
region and at the NE side, while only 3-$\sigma$ upper limits are obtained at the
SW side. Such result indicates that the shock at the NE side is stronger than 
that at the SW side. This is consistent with \cite{Garcia-Burillo2001}, who
found extended off-nuclear SiO $J$=2-1 emission on the NE side, indicating
strong molecular shock on large scales. On the other hand, \cite{Lester1990,
Mouri1989} found higher ratio of H$_2$ 1-0 $S$(1)/Br$\gamma$ at the NE side
than that at the SW side, consistent with the scenario that the NE shock is
stronger.  However, the linewidths of HCN, HCO$^+$ and HNC do not show obvious
differences at both sides, indicating that the shocks may not heavily broaden the
linewidth and influence the global properties of dense gas. 

\subsection{HCO $J$=1-0}

As shown on the right panel of Figure \ref{fig:H13CN_H13CO+}, HCO $J$=1-0, as a
good tracer of PDR \citep{Garcia-Burillo2002,
Gerin2009, Martin2009b}, is stronger at the SW side than that at the NE side.
In addition, \cite{Garcia-Burillo2002} also found a giant PDR of $\sim$650 pc
size in M$\,$82 with HCO 1-0 mapping observation, using the IRAM Plateau de Bure
Interferometer. The result suggests that the chemistry of the SW molecular side
is dominated by the PDR, which shows typical features of an evolved starburst
\citep{Aladro2011b, Fuente2008}.

\subsection{H$_2$CO $J$=2-1}

With high critical density of 1.6$\times$10$^6$ cm$^{-3}$ \citep{Kennicutt2012}, the H$_2$CO $J$=2-1 transition can also trace
dense gas \citep{Bayet2008}, which traces the high-excitation component of the
molecular gas in M$\,$82 very well \citep{Muhle2007}. The H$_2$CO 2-1 emission at
the SW side is stronger than that at the NE side, indicating higher excitation
conditions at the SW side.  Such difference may be caused by the asymmetric
outflow \citep{Shopbell1998}, which may also impact the NE disk
\citep{Seaquist2001, Veilleux2009}. 

\subsection{HC$_3$N $J$=10-9}

As a warm and dense gas tracer\citep{Tanaka2018}, HC$_3$N can be easily
destroyed by UV radiation and Cosmic Rays \citep{Rico2020,Costagliola2010}.
Therefore, HC$_3$N is mainly excited by collision. HC$_3$N lines are 
bright enough to be detected in the local galaxies NGC 4418, NGC 253, IC 342 NGC 6240
and so on \citep{Costagliola2011,Aladro2011a,Li2019}. These detections all
indicate that the optical depth of HC$_3$N is very small ($\tau\ll1$) in most
cases, because it has many energy populations. 

Assuming similar excitation temperature, the abundance ratio of
$I$(H$^{13}$CN)/$I$(HC$_3$N) can be estimated by their intensity ratio.  This
ratio does not show large variation among different positions,  indicating that
the distribution of H$^{13}$CN and HC$_3$N might be uniform along the major
axis. The results also implied that the HC$_3$N emission is either spatially
separated from PDRs, or the PDRs might be weak, because UV photons from PDRs
can dissociate HC$_3$N.


\end{document}